\documentclass[pra, reprint, twocolumn, superscriptaddress, amsfonts, amssymb, amsmath]{revtex4-2}
\usepackage[english]{babel}

\usepackage{graphicx}

\graphicspath{{./Images/},{./ImagesAppendix/},
{./images/},{./imagesAppendix/}}

\usepackage{physics}
\usepackage{bm}
\usepackage{braket}
\usepackage{pgfplots}
\usepackage{array}
\usepackage{makecell}
\usepackage{amsmath}
\usepackage{tikz}
\usetikzlibrary{positioning, arrows.meta}
\usepackage{subfigure}
\usepackage{commath}
\usepackage{graphicx,bm}
\usepackage{verbatim}
\usepackage{flafter}
\usepackage{float}
\usepackage{varioref}

\usepackage{mathtools}
\DeclarePairedDelimiter\ceil{\lceil}{\rceil}

\usepackage{color}
\usepackage[colorlinks,citecolor=darkBlue,linkcolor=darkBlue,
urlcolor=blue,hyperindex]{hyperref}
\definecolor{darkBlue}{rgb}{0.08, 0.13, 0.4}

\definecolor{THc}{rgb}{0.9,0.3,0.2}
 
\begin{document}


\title{Coherent excitation transport through  ring-shaped networks}
\author{Francesco Perciavalle}
\affiliation{Quantum Research Center, Technology Innovation Institute, P.O. Box 9639 Abu Dhabi, UAE}
\affiliation{Dipartimento di Fisica dell’Universit\`a di Pisa and INFN, Largo Pontecorvo 3, I-56127 Pisa, Italy}

\author{Oliver Morsch}  
\affiliation{CNR-INO and Dipartimento di Fisica dell’Universit\`a di Pisa, Largo Pontecorvo 3, 56127 Pisa, Italy}

\author{Davide Rossini}
\affiliation{Dipartimento di Fisica dell’Universit\`a di Pisa and INFN, Largo Pontecorvo 3, I-56127 Pisa, Italy}

\author{Luigi Amico}
\affiliation{Quantum Research Center, Technology Innovation Institute, P.O. Box 9639 Abu Dhabi, UAE}
\address{Dipartimento di Fisica e Astronomia, Via S. Sofia 64, 95123 Catania, Italy}
\affiliation{INFN-Sezione di Catania, Via S. Sofia 64, 95127 Catania, Italy}
\affiliation{Centre for Quantum Technologies, National University of Singapore 117543, Singapore}

\date{\today}

\begin{abstract}
The coherent quantum transport of matter wave through a ring-shaped circuit attached to leads  defines an iconic system in mesoscopic physics that has allowed both to explore fundamental questions in quantum science and to draw important avenues for conceiving devices of practical use.
Here we study the source-to-drain transport of excitations going through a ring-network, without propagation of matter waves. We model the circuit in terms of a spin system with specific long-range interactions that are relevant for quantum technology, such as Rydberg atoms trapped in optical tweezers or ion traps. Inspired by the logic of rf- and dc-SQUIDs, we consider rings with one and  two local energy offsets, or detunings. As a combination of specific phase shifts in going though the localized detunings and as a result of coherent tunneling, we demonstrate how the transport of excitations can be controlled, with a distinctive dependence on the range of interactions. 
\end{abstract}

\maketitle

\section{Introduction}
Quantum transport in mesoscopic circuits deals with matter propagating in networks  characterized  by a spatial scale comparable with the particles coherence length~\cite{beenakker1991quantum,blanter2000shot}. In this regime, quantum effects as quantum tunneling, conductance quantization, flux-quantization, Aharanov-Bohm effect, etc, play a prominent role~\cite{datta2005quantum,nazarov2009quantum,stone1992theory}. Recently, transport properties of cold atoms guided in versatile and flexible laser-generated circuits have been studied both theoretically and experimentally~\cite{chien2015quantum,stadler2012observing, husmann2015connecting,corman2019quantized,haug2019aharonov, haug2019topological, haug2019andreev, lau2023atomtronic}. In fact, through widely tunable interactions and disorder, new paradigms in mesoscopic physics have been defined, both for bosonic and for fermionic systems, with a great potential for basic quantum science and applications~\cite{amico2021roadmap,amico2022colloquium}. 

In this paper we refer to one of the most iconic systems of mesoscopic physics that has led to far reaching implications: a mesoscopic ring-shape track connected to source and drain leads~\cite{beenakker1991quantum,blanter2000shot,buttiker1984quantum,gefen1984quantum,washburn1986aharonov}. The tunneling through scattering impurities or localized barriers placed in the ring circuit can induce specific phase shifts in the particles wave function~\cite{griffiths2018introduction}. In this way, the source-to-drain current can be controlled, a fact that is relevant both to study fundamental features of quantum interference and to engineer mesoscopic quantum devices with enhanced performances. 
With a similar logic, neutral bosonic matter-wave current oscillations have been predicted and analysed in Ref.~\cite{haug2019aharonov}. Rings with one or two localized barriers can define the bosonic analog of rf- and dc-SQUIDs~\cite{ryu2013experimental,ryu2020quantum,krzyzanowska2023matter,amico2022colloquium}. Such cold-atom implementations pave the way to rotation sensors based on the Sagnac effect~\cite{barrett2014sagnac}. 

Here  we study the source-to-drain quantum transport through a ring-shaped circuit, in which the dynamics occur in terms of excitations, rather than of matter. The implementation we rely on is a circuit made of Rydberg atoms trapped in tweezers or ions trapped in suitable electromagnetic fields~\cite{blatt2012quantum,monroe2021programmable,browaeys2020many, bernien2017probing}. Indeed, in such systems the motion of the atoms or of the ions can be neglected, with the relevant dynamics occurring as the transfer of excitations of internal energy states. Moreover, both Rydberg atoms and ions can be trapped in a large variety of different geometries and with a remarkable control of the physical conditions~\cite{barredo2016atom,schymik2020enhanced,birkl1992multiple,kiesenhofer2023controlling}.
In contrast with the cases considered so far, here we deal with systems with a long-range interaction, capturing the characteristic physics of Rydberg atoms and trapped ions~\cite{maier2019environment,barredo2015coherent,yang2019quantum}.

We implement a specific quench dynamics that we demonstrate to lead to the sought drain-to-source excitation transport: after initializing the excitation in the source, it propagates along the two arms of the ring and, after many scattering events determined by the interaction, reaches the drain. Dealing with a closed system, the excitations population in the source and in the drain displays characteristic oscillations.        
We shall demonstrate that the entire excitation dynamics and then the source-to-drain transport can be controlled by suitable energy level detunings localized in the ring track. Such detunings play a similar role of what the aforementioned local potential barriers do for matter waves~\cite{haug2019aharonov}. Indeed the presence of detunings gives a non trivial time-dependent phase to the time-evolved state. Inspired by the rf-and dc-SQUID concepts, we address the cases of one and two localized detunings. 
\begin{figure*}[!t]
\centering
\includegraphics[width=0.75\textwidth]{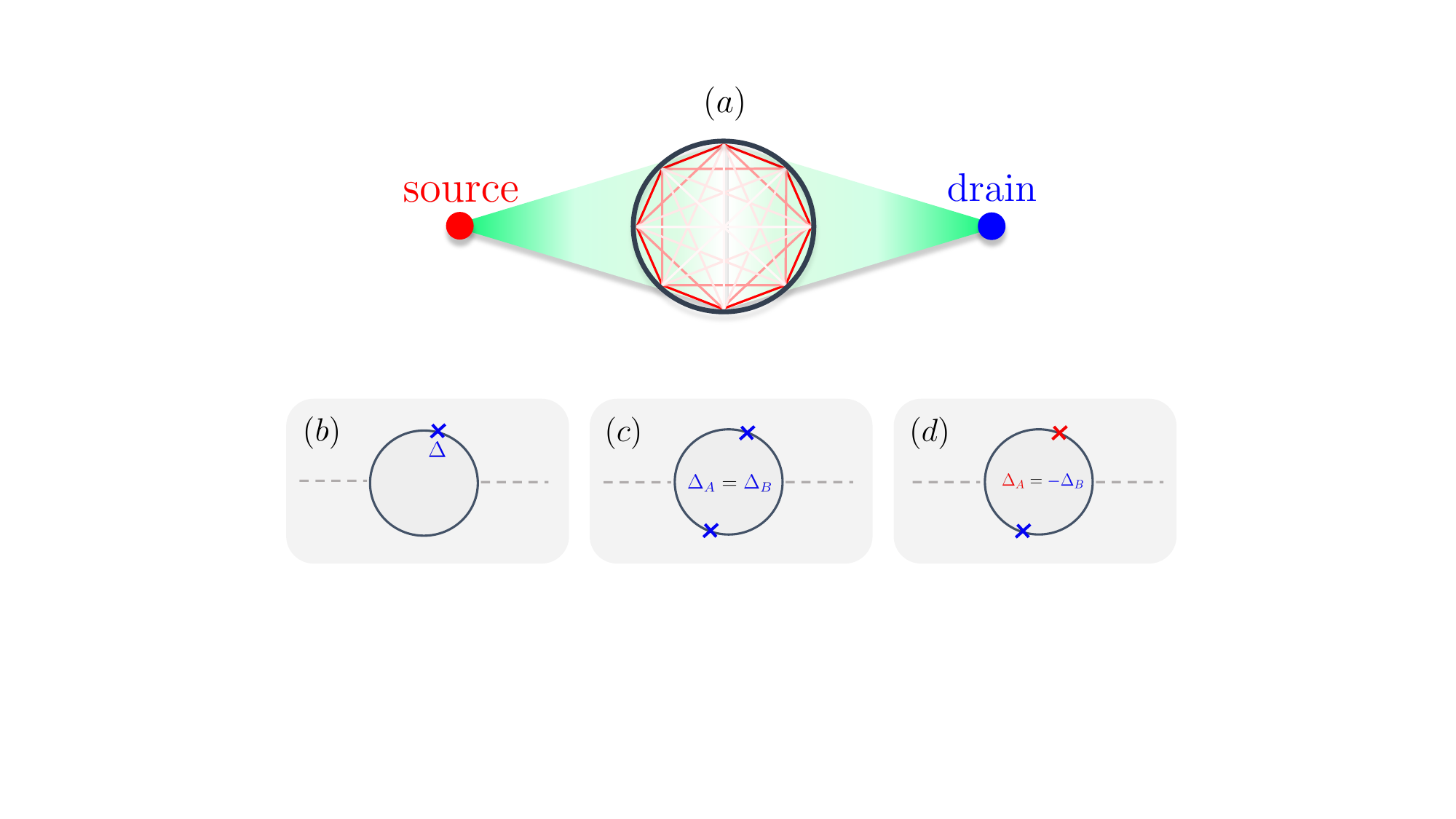}
\caption{$(a)$ Pictorial representation of the system: two single-site leads, a source and a drain, are connected to a ring lattice. The various sites are connected with a hopping strength that scales as $1/d^{\alpha}$, $d$ being the distance between sites. Boxes show three possible configurations we are going to study: $(b)$ one barrier, $(c)$ two barriers with the same sign, $(d)$ two barriers with opposite sign. The number $N_r$ of sites in the ring lattice is such that $N_r$ is even, while $N_r/2$ is odd. The asymmetric location of the detunings with respect to the leads is chosen to guarantee different paths for the excitation on the two arms.}
\label{fig:scheme}
\end{figure*}
The paper is organized as follows: in Sec.~\ref{sec:Model}, we introduce the model and the possible experimental platforms on which it can be realized. In Sec.~\ref{sec:sing_det}, we study the dynamics of the leads and the ring population in presence of a single detuning in one arm of the ring. In Sec.~\ref{sec:double}, we perform the same type of simulations but in presence of two detunings, one for each arm of the ring, with the same height and the same sign first, with opposite sign secondly. Our conclusions are drawn in Sec.~\ref{sec:concl}. The Appendix provides a more detailed analysis of the model properties and of the mechanisms that lead to the peculiar dynamics observed in the  main text.

\section{Model and Methods}
\label{sec:Model}
We consider a system comprised of two leads, the source ($\mathcal{S}$) and the drain ($\mathcal{D}$), each of them modeled by a single site containing at most one excitation, connected to a ring network ($\mathcal{R}$) of $N_r$ sites. Localized detunings, or impurities ($\mathcal I$), are placed in the ring lattice and arranged in different configurations (see Fig.~\ref{fig:scheme}). The excitation in each site is modelled as a two-level atom system $\{\ket{\uparrow}, \ket{\downarrow} \}$. 
The system Hamiltonian reads
\begin{equation}
\hat{\mathcal{H}}=\hat{\mathcal{H}}_{\mathcal{R}}+\sum_{\mathcal{L}=\mathcal{S},\mathcal{D}}\hat{\mathcal{H}}_{\mathcal{L}\mathcal{R}}+\hat{\mathcal{H}}_{\mathcal{S}\mathcal{D}},
\end{equation}
with
\begin{subequations}
\begin{align}
& \hat{\mathcal{H}}_{\mathcal{R}} = \sum_{i < j\in \mathcal{R}}\frac{g}{d_{i,j}^\alpha}\left(\hat{\sigma}_i^+\hat{\sigma}_j^- + \rm H.c.\right)+\hat{\mathcal{H}}_{\rm Det},\label{eq:Ring}\\
& \hat{\mathcal{H}}_{\mathcal{L}\mathcal{R}} = \sum_{j\in \mathcal{R}}\frac{g}{d_{\mathcal{L},j}^\alpha}\left(\hat{\sigma}_j^+\hat{\sigma}_{\mathcal{L}}^- + \rm H.c.\right), \quad (\mathcal{L}=\mathcal{S}, \mathcal{D}),\label{eq:LR}\\
& \hat{\mathcal{H}}_{\mathcal{S}\mathcal{D}} = \frac{g}{d_{\mathcal{S},\mathcal{D}}^\alpha}\left(\hat{\sigma}_{\mathcal{S}}^+\hat{\sigma}_{\mathcal{D}}^- + \rm H.c.\right),\label{eq:SD}\\
& \hat{\mathcal{H}}_{\rm Det} = \sum_{j\in \mathcal{R}} \sum_{j_l\in  \mathcal{I}} \Delta_{j}\hat{n}_{j}\delta_{j,j_l}.
\label{eq:Ham_1imp}
\end{align}
\end{subequations}
Here $\hat{\sigma}^a_\ell$ ($a=x,y,z$) denote the spin-1/2 Pauli matrices on a given atom and $\hat{\sigma}^\pm_\ell = \tfrac12 (\hat{\sigma}^x_\ell \pm i \hat{\sigma}^y_\ell)$ are the corresponding raising/lowering operators,
with $\ell = \{ \mathcal{S}, \mathcal{D}, 1, \ldots, N_r \}$ labeling one of the atoms in the leads or in the ring. 
The Hamiltonian part~\eqref{eq:Ring} describes the intra-ring hopping in the presence of detunings~\eqref{eq:Ham_1imp}, while Eq.~\eqref{eq:LR} describes the leads-ring hopping and Eq.~\eqref{eq:SD} describes the direct source-drain hopping. 
We suppose that the excitation $\ket{\uparrow}$ can hop from site to site with a strength scaling as $1/d^{\alpha}$, where $d$ denotes the distance between the sites and $\alpha \geq 0$ is a parameter inversely related to the hopping range. In particular, $d_{i,j}$ is the intra-ring distance ($i,j = 1,\ldots N_r$), $d_{\mathcal{S},j}$ and $d_{\mathcal{D},j}$ are the distances between the leads and all the sites of the ring, while $d_{\mathcal{S},\mathcal{D}}$ is the distance between source and drain (see \ref{App:Dist}).
Hereafter we mainly concentrate on the weak lead-ring coupling limit, corresponding to $K_{nn}\ll J_{nn}$, where $K_{nn}$ and $J_{nn}$ are the leads-ring and intra-ring nearest-neighbor hopping strengths, respectively (see \ref{App:Dist} for details). We also consider an even number $N_r$ of atoms in the ring, in such a way that $N_r/2$ is odd.

We should stress that we are interested in the cases $1 \le \alpha\le 3$, where the nearest-neighbor and the fully connected models correspond to $\alpha\rightarrow\infty$ and $\alpha=0$, respectively.  Trapped ions in linear chains provide a powerful platform to realize quantum spin Hamiltonians in which the spin-spin interaction or the hopping strength scale approximatively as $d^{-\alpha}$, where $\alpha \in (0, 3)$~\cite{blatt2012quantum, arrazola2016digital, richerme2014non, maier2019environment,bohnet2016quantum,brown2016co}; the $\ket{\downarrow}$ and $\ket{\uparrow}$ states are two internal electronic states of the ions. The implementation of different geometries is more challenging, but two-dimensional spin Hamiltonians have been realized as well~\cite{rajabi2019dynamical,duca2023orientational,britton2012engineered, bohnet2016quantum, yoshimura2015creation, richerme2016two}. Concerning rings, circular traps of ions have been realized as well~\cite{birkl1992multiple, noguchi2014aharonov,li2017realization}. 
The case $\alpha=3$ can be also implemented in Rydberg atoms. For such systems, $\ket{\downarrow}$ and $\ket{\uparrow}$ are two Rydberg states of opposite parity (i.e., $\ket{S}$ and $\ket{P}$)~\cite{browaeys2020many, barredo2015coherent, chen2023continuous, bornet2023scalable}. Thanks to the high control on the geometry in which the atoms are located in these systems~\cite{barredo2016atom, schymik2020enhanced, endres2016atom}, it is possible to realize ring-shaped and more complicated geometries~\cite{lienhard2020realization}. 

We focus on three possible configurations of the localized detunings in the ring lattice. The case $\mathcal{I}=\{ j_0\}$ corresponds to one localized detuning (Fig.~\ref{fig:scheme}b), the path crossed by the excitation in the two arms of the ring is clearly different. The two other cases correspond to $\mathcal{I}=\{ j_A,j_B\}$ with $j_A$ and $j_B $ in diametrical position, $\Delta_A=\Delta_B$ and $\Delta_A=-\Delta_B$ respectively (Fig.~\ref{fig:scheme}c, d). Since $d_{\mathcal{S},j_A}\neq d_{\mathcal{S},j_B}$, also in this case, the path crossed by the excitation fraction in one arm of the ring is different from the other. The local detunings are experimentally feasible through local AC-Stark shifts with a focused laser on the site of interest~\cite{maier2019environment,de2017optical}.

To study the excitation transport, we apply the following quench protocol. At $t=0$ we localize a single excitation in the source:
\begin{equation}
\ket{\psi(0)}=\ket{\uparrow}_{\mathcal{S}}\otimes\ket{\downarrow,\ldots,\downarrow}_{\mathcal{R}}\otimes \ket{\downarrow}_{\mathcal{D}},
\label{eq:instate}
\end{equation}
then we evolve the state via the Schrödinger equation $\ket{\psi(t)}=\exp \big(-i\hat{\mathcal{H} t} \big) \ket{\psi(0)}$. 
To monitor the dynamics, we scrutinize the number of excitations in the source, the ring, and the drain:
\begin{equation}
    \hat{n}_{\mathcal{R}}=\dfrac{1}{2} \sum_{j\in \mathcal{R}}(\hat{1}_{j}+\hat{\sigma}^z_{j});
    \quad
    \hat{n}_{\mathcal{L}}=\dfrac{1}{2} (\hat{1}_{\mathcal{L}}+\hat{\sigma}^z_{\mathcal{L}}),
    \quad
    (\mathcal{L}=\mathcal{S},\mathcal{D}).
\end{equation}
Since the Hamiltonian conserves the total number of excitations $\hat{n}_{tot}=\hat{n}_{\mathcal{S}}+\hat{n}_{\mathcal{R}}+\hat{n}_{\mathcal{D}}$,
with the initial condition~\eqref{eq:instate}, we can work in the one-excitation sector, namely $\braket{\psi(t)|\hat{n}_{tot}|\psi(t)}=1$.
Thus, we define the projector $\hat{P}$ represented by a $N\times 2^{N}$ matrix that projects operators and states of the full Hilbert space into operators and states of the one excitation sector. The states and operators with which we will work are of the form $\hat{P}\ket{\psi}$ and $\hat{P}\hat{O}\hat{P}^{\dagger}$, $\ket{\psi}$ and $\hat{O}$ being a generic state and operator acting on the full Hilbert space.

\section{Dynamics in the presence of a single detuning}
\label{sec:sing_det}

\begin{figure*}[!t]
\centering
\includegraphics[width=0.85\textwidth]{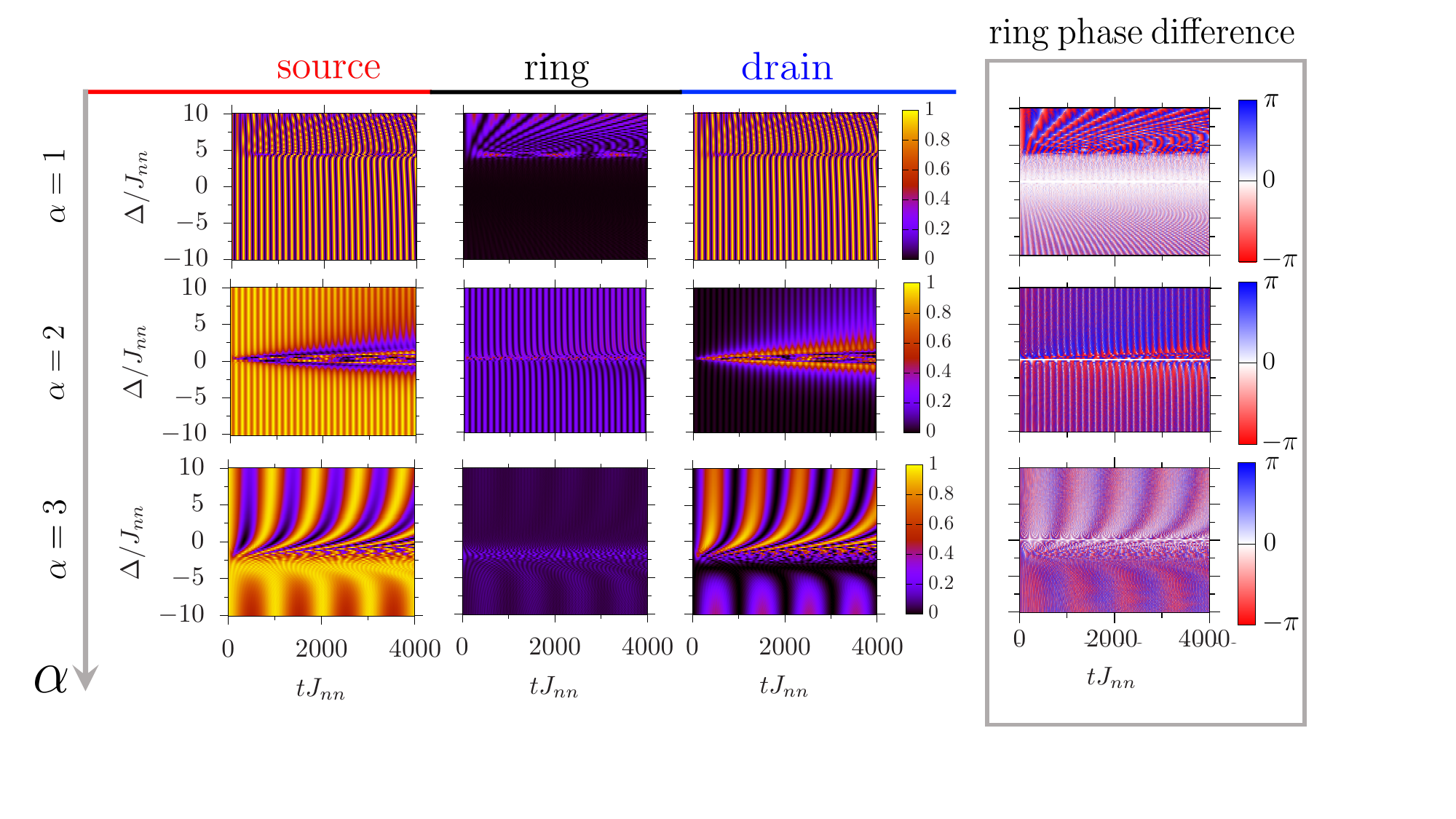}
\caption{Central-left panels: excitation populations in the source $P_{\mathcal{S}}$, the ring $P_{\mathcal{R}}$, and the drain $P_{\mathcal{D}}$, for three different values of the hopping range $\alpha=1,2,3$ from top to bottom. Right panels: ring phase difference $\delta\phi$ between two opposite sites located in the two arms of the ring, for the three values of $\alpha$ considered. Excitation populations and phase difference are plotted in function of time ($x$-axis) and detuning ($y$-axis) of the single barrier located in the ring at position $j_0=\ceil{N_r/4}$. The population values are indicated by the color bar that goes from $0$ to $1$, the phase difference by the one that goes from $-\pi$ to $\pi$. The number of sites in the ring is $N_r=14$. The leads-ring coupling is $K_{nn}=J_{nn}/10$.}
\label{fig:3dplots_singbarr}
\end{figure*}

Let us first consider the case of a single localized detuning (Fig.~\ref{fig:scheme}b). 
We analyze the dynamics of the number of excitations in the source, the ring, and the drain, as well as the phase difference behavior between the two arms of the ring (in particular, the phase difference between two opposite sites located in the two different arms).
Numerical results are summarized in Fig.~\ref{fig:3dplots_singbarr}.

\subsection{Excitation dynamics in presence of a single detuning}
Because of the weak coupling assumption, insight on the general features of the excitation dynamics can be achieved by looking at the (discrete) energy spectrum of the leads and the ring separately. The initial state $\ket{\psi(0)} \equiv \ket{\mathcal{S}}$ is assumed to be at zero energy. The ring Hamiltonian in the single-particle sector has $N_r$ eigenstates with non-zero energies; they result to be doubly degenerate except for the ground state and the highest excited state. The role of the localized detuning is to break the degeneracies and shift the ring levels.  For specific values of the detuning  $\Delta_{res}$ the ring energy levels can be resonant with the initial source state $\ket{\mathcal{S}}$ (see \ref{app:Ring_shift}). In this case, the excitation is transferred  from source to the drain and backward to the source, 
through the ring that is well populated (see Fig.~\ref{fig:3dplots_singbarr}, central-left panels). Since the total number of excitations is conserved and the ring is populated, during its time evolution the population is distributed along the leads and the channel. Far from resonance, coherent source-drain oscillations display a marked dependence on $\Delta$ as soon as $\alpha$ increases. This means that the source-drain dynamics depends on the characteristics of the ring and not only on the source-drain direct coupling $\hat{\mathcal{H}}_{\mathcal{S}\mathcal{D}}$. 
In this regime, the transport can occur in terms of cotunneling processes~\cite{nazarov2009quantum, averin1990virtual, tran2008sequential}: Similarly to the transport occurring in quantum dots in the regime in which sequential tunneling cannot happen (because of the Coulomb blockade), a source-to-drain transport in our system can occur through virtual transitions in the ring energy levels; in this regime, the ring is found with a low population of excitations.

Due to the presence of more than nearest-neighbor hopping processes, specific features of the system emerge when leads and ring-shape track are not treated separately (see Fig.~\ref{fig:scheme}). 
Specifically, because of a combination of  geometrical effects and energy level configuration, we found that the transport  displays a marked dependence on $\alpha$, especially in the (leads-ring) nearly resonant regimes. 
For $\alpha=1$, the source and drain populations oscillate regularly until $\Delta/J_{nn}\approx 5$, value for which the rings starts to be populated and the dynamics inside the leads is less regular. Far from that value, the ring is not populated and the source-drain oscillations do not depend on the detuning.
For $\alpha=2$, the dynamics is quite different: in the ($\Delta$,$t$) plane, a V-pattern appears. 
On the borders of the V-pattern, the drain and source populations are respectively maximum and minimum. The vertex of the V is in correspondence of $\Delta_{res}$, where the ring is more populated. Another important feature of this regime is the population dynamics for values of detuning far from the resonance. The source is at every time the most populated, the drain the less populated. The ring population is small but not zero, it shows a weak oscillation with those of the source. 

The $\alpha=3$ regime is characterized by slow (compared with the previous cases $\alpha=1$ and $\alpha=2$) coherent transfer of excitation from the source to the drain interrupted at the value of $\Delta$ for which source and ring are resonant. As a specific feature of this case, we note a localization of the excitation in the source occurring for values of $\Delta$ slightly smaller of $\Delta_{res}$. Indeed, such localized source state appears in the analysis of the full leads-ring system Hamiltonian spectrum (see~\ref{app:allspectrum}). For different values of $\Delta$, the complete states have also a substantial projection in the drain state and therefore the dynamics  results to display the characteristic  source-drain coherent oscillations. 

\subsection{Phase difference dynamics in presence of a single detuning}

We now analyse the phase of the excitations flow between the two arms of the ring. To this end, we compute the phase difference between the two symmetric sites of the arms, labeled by $i=N_r/2-1$ and $j=N_r/2+1$. The argument of the two-body correlator $\braket{\hat{\sigma}_i^{+}\hat{\sigma}_j^-}$ gives the phase difference $\delta \phi$ between the two sites (see \ref{app:phase_diff} for details). To corroborate our results, we analytically studied a straight finite chain and found that the phases of the coefficients of the time evolved state in the position basis depend non trivially on $\Delta$. 

\begin{figure*}[!t]
\centering
\includegraphics[width=0.85\textwidth]{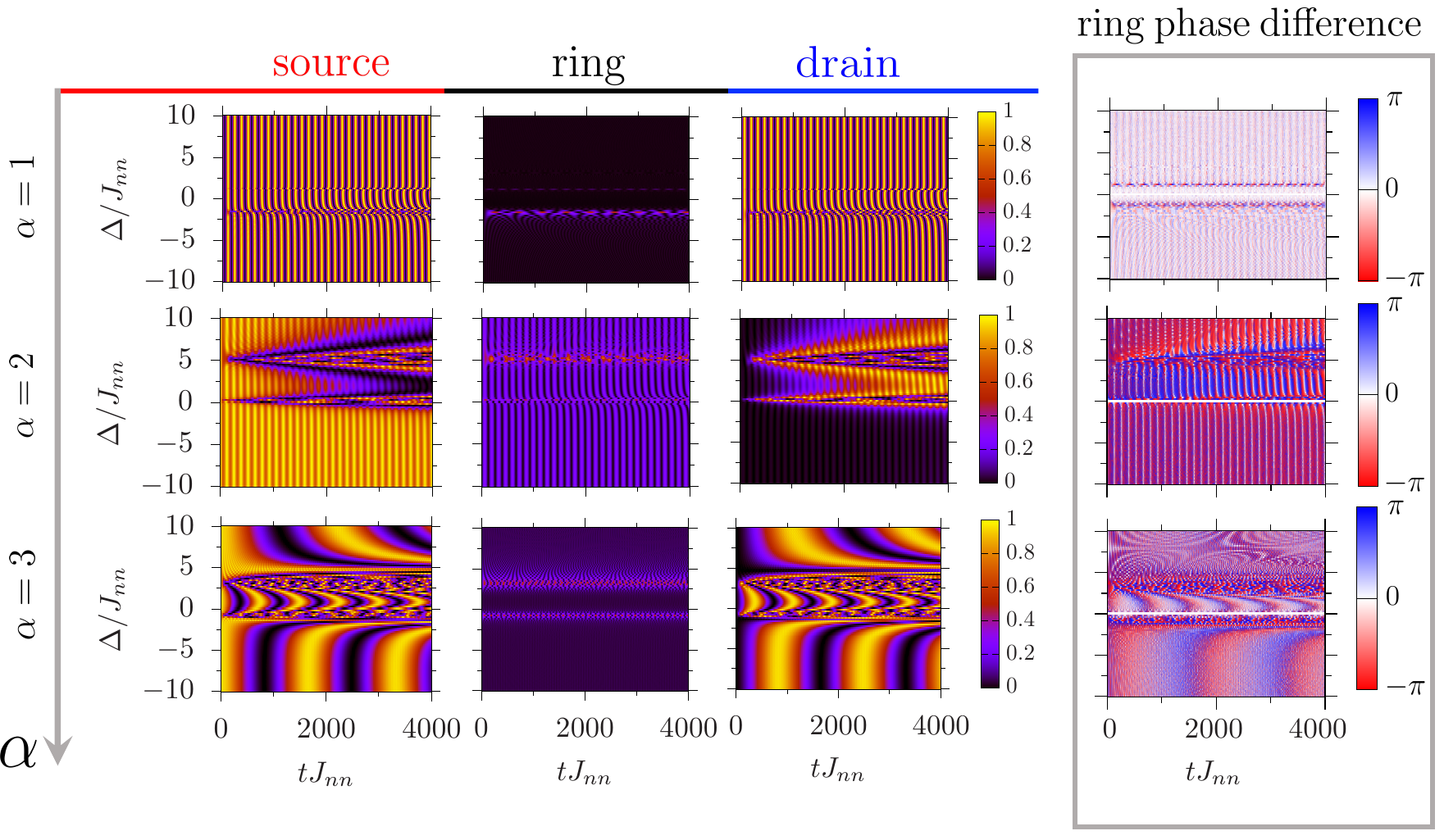}
\caption{Populations and ring phase difference as in Fig.\ref{fig:3dplots_singbarr}, for two equal detunings ($\Delta_A=\Delta_B=\Delta$) located at $j_A=\ceil{N_r/4}$ and $j_B=\ceil{3N_r/4}$. The Hamiltonian parameters are set as in Fig.\ref{fig:3dplots_singbarr}.}
\label{fig:3dplots_2eqbarr}
\end{figure*}

The rightmost panels of Fig.~\ref{fig:3dplots_singbarr} report the phase difference as a function of time and detuning, for different values of $\alpha$. Our results indicate that, as for matter waves with a localized potential barrier, the localized detuning causes a phase shift between the two arms of the ring. The dependence of the phase difference on the detuning is evident for each value of $\alpha$. In particular, it follows the ring population oscillation frequency. However, also when the ring population oscillation amplitude is small and hardly observable, the phase difference can present non negligible oscillations. For instance, for $\alpha=1$ and negative values of the detuning, there are evident phase difference oscillations while the ring is basically not populated. In a window around $\Delta=0$, the phase difference is zero because the paths on the two arms are the same. On the other hand, in correspondence of the resonances, as soon as the ring becomes populated, we observe strong oscillations in the phase difference.
In general, the phase oscillation follows the population oscillation, independently on the amplitude of the latter. If there is a little amount of excitation that moves in the ring, the phase difference oscillates with an amplitude that is significantly different from zero.

\section{Dynamics in presence of double detuning}
\label{sec:double}

In this section we focus on the source-to-drain dynamics of the number of excitations in the case in which two atoms in the ring are detuned. We first study the case in which the two detunings have the same sign (Fig.~\ref{fig:scheme}c) and then the case in which they have same magnitude but opposite sign (Fig.~\ref{fig:scheme}d).

\subsection{Excitation dynamics in presence of double detunings with the same sign}
\label{DoubleSameSign}
Also in this case, the hopping range has pronounced effects on the dynamics. The combined energy shifts caused by the two detunings and hopping range  give rise to a resonant transport occurring in correspondence to two resonances (instead of the one in the previous single detuning case) that turns out to be not symmetric with respect to $\Delta=0$. 
The nature of the level shift that brings to resonance can be accessed analytically and numerically looking at the ring spectrum respectively in the nearest-neighbor and long-range cases (see~\ref{app:Ring_shift}).  The population dynamics is reported in the central-left panels of Fig.~\ref{fig:3dplots_2eqbarr}.

For $\alpha=1$, the dynamics is characterized by fast oscillations between source and drain with a frequency that is nearly independent of detuning. The ring is almost never populated except for the two $\Delta_{res}$ values, where the dynamics in the leads result to be erratic.
In the case $\alpha=2$, the dynamics clearly changes. Far from resonances, the dynamics result to be  similar to the  one observed in the single detuning  case. In this case we see two V-shaped  patterns in the $(\Delta,t)$ plane around the two resonances. We notice a depletion in the source and a filling of the drain in correspondence of the meeting points of the  two V-shaped  patterns.
Finally, as in the case with single detuning, for $\alpha=3$ the transport between source and drain is slower and more affected by the process occurring in the ring track. In particular, tuning $\Delta$ the transport in the system shows markedly different features. For large (positive or negative) values of $\Delta$, the  transport of excitations is slow-down, with  the ring resulting weakly populated. Around the two resonant levels the transport becomes erratic and the ring results populated. In between the two resonances, we observe regular and fast oscillations of the population in the leads. 

In \ref{app:strong_coupling} we report strong leads-ring coupling results for $\alpha=3$. It is observed that increasing $K_{nn}/J_{nn}$ the source-drain oscillations weaken. The latter survive until $K_{nn}$ remains smaller than $J_{nn}$; when the couplings are comparable, the leads are weakly populated and clear oscillations are no longer observed. In this scheme, the spectrum is not divided into leads and ring states far from the resonance, so the ring is populated independently of $\Delta_{res}$.

\subsection{Excitation dynamics in presence of double detunings with opposite sign}

\begin{figure*}[!t]
\centering
\includegraphics[width=0.85\textwidth]{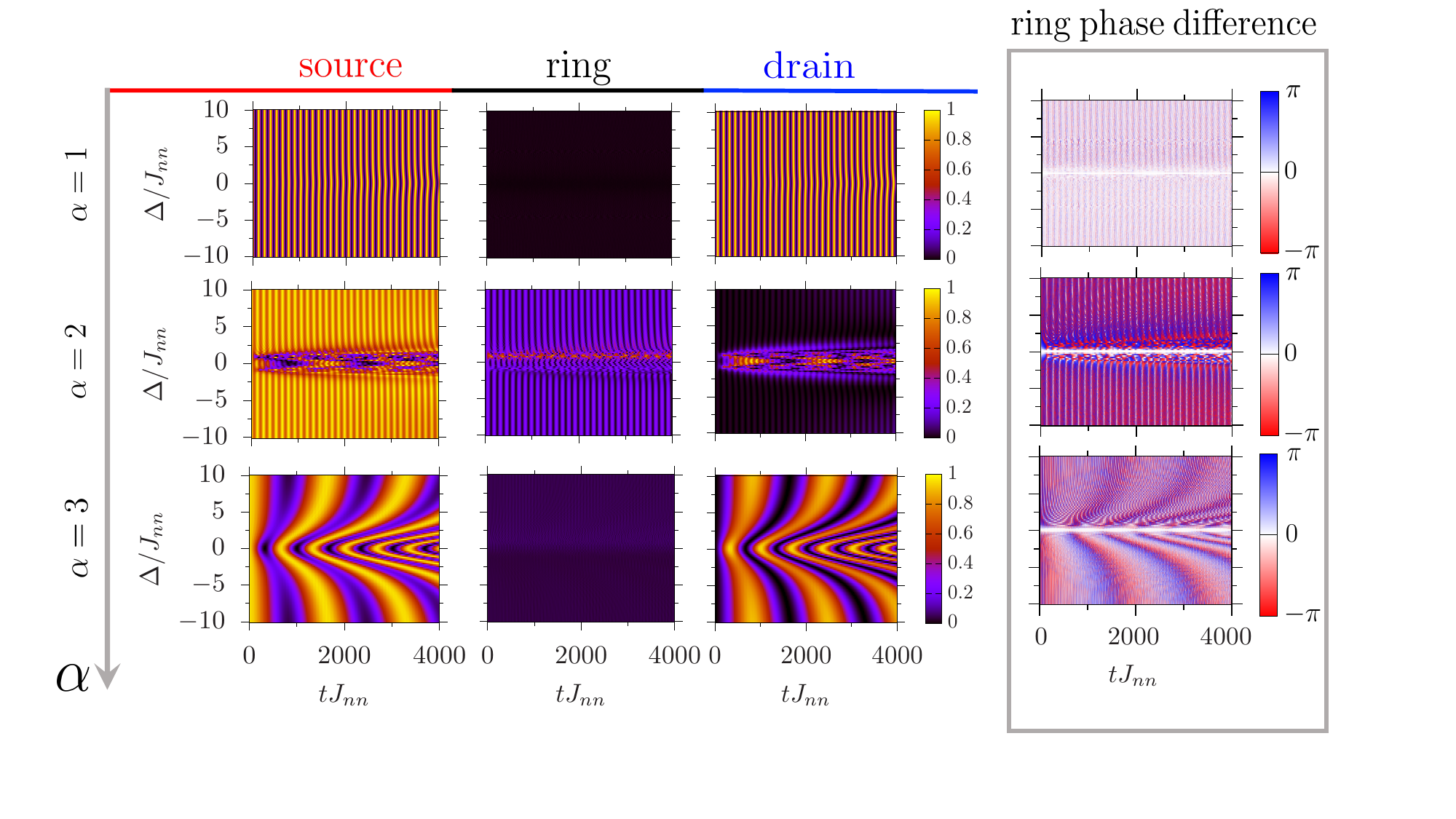}
\caption{
Same as in Fig.~\ref{fig:3dplots_2eqbarr}, but for two detunings with equal magnitude but opposite sign: $\Delta_A=-\Delta_B=\Delta$.}
\label{fig:3dplots_2opbarr}
\end{figure*}

We complete our analysis by studying the configuration of double equal barrier with opposite sign. Also here, the three different regimes can be observed by changing $\alpha$. The double barrier regime is basically different from the other two because the first perturbative order energy correction is zero (see \ref{app:Ring_shift}), thus it is harder to get resonances.
Results for the population dynamics are reported in the central-left panels of Fig.~\ref{fig:3dplots_2opbarr}.
For $\alpha=1$, the dynamics is characterized by fast oscillations between source and drain, the ring is never resonant with initial state and so is never populated. 
For $\alpha=2$, the ring spectrum presents two symmetric resonances (instead of the asymmetric ones of the previous subsection)  with the zero energy mode. This feature is corroborated by a perturbative analysis showing that  the resonances depends as $\Delta^2$ (see~\ref{app:Ring_shift}).
For large (positive and negative) values of $\Delta$, the population in the drain is minimal, while a small fraction of excitation oscillates between source and ring. 
For values of $\Delta$ between the two resonances, a source-drain slow oscillation given by the recombination of the two V-shaped patterns appear.
The regime corresponding to $\alpha=3$ is characterized by slow coherent oscillations between the two leads, the ring track being weakly populated. Differently from the $\alpha=1$ case, the value of $\Delta$ determines the oscillation frequency between source and drain. We found that the excitation transfer slows down by increasing $\Delta$. 
\subsection{Two arms phase difference with double detuning}
The rightmost panels of Figs.~\ref{fig:3dplots_2eqbarr} and~\ref{fig:3dplots_2opbarr} report the phase difference between the sites of the two arms closest to the drain in presence of two detunings. As in the case of single detuning, the phase difference oscillations are minima in amplitude for $\alpha=1$ far from resonances. For $\alpha=2$ we observe strong phase difference oscillations, evident oscillations are observed also for $\alpha=3$. In all the cases analyzed so far, the phase difference oscillations follow in frequency those of populations. Also in this case, the phase oscillations are correlated to the ring population oscillations: the phase oscillation amplitude is bigger when the ring is more populated. However, a little amount of excitation in the ring is sufficient to have substantial oscillations in the phase.

In~\ref{app:phase_diff} we also discuss the phase difference dynamics for different locations of the detunings. As expected, phase oscillations strongly depend on it; if the detunings are located in such a way that the excitation fractions follow the same paths on the two arms of the ring, the phase difference is zero for any $\Delta$. This result can be achieved putting two equal detunings in two diametrically opposite zones of the arms. Otherwise, if the resonance location in $\Delta$ is the same and the two paths are different, the dynamics differs with the one considered here with specific features depending on the geometry.

\section{Discussion}
\label{sec:concl}

We studied the source-drain excitation transport through a ring lattice track in which, in one or two sites, the atoms are detuned by $\Delta$ respect to the others; such detunings act as local energy shifts in the system and therefore play the role of localized impurities. 
Motivated by the know-how in quantum technology, the system is described by a XY Hamiltonian with  $1/d^{\alpha}$ hopping; specific values of $\alpha$ that are relevant for ion traps~\cite{monroe2021programmable,arrazola2016digital,jurcevic2014quasiparticle} and Rydberg atoms~\cite{browaeys2020many} localized in optical tweezers are considered. This feature should be contrasted with previous studies corresponding to nearest-neighbor hopping~\cite{haug2019aharonov}.
As detailed below, the transport can be controlled by tuning $\Delta$, with distinctive features of $\alpha$.

The energy excitation is initialized in the source lead, with a zero energy state. Then it is transferred to the ring by quenching the interaction with the rest of the system. In the weak leads-ring tunneling regime,  we note that, once the excitations are transferred  to the ring track,  they propagate within the ring with a fast time dynamics.
The presence of localized detunings causes a shift of the energy levels of the XY Hamiltonian of the ring and, for specific values of $\Delta$, a state resonant with the source one can occur.   
Inspired by rf- and dc-SQUIDs, we addressed the two cases of one and two detunings; in turn, we considered two detunings with the same or with opposite sign. 

Resonant and non-resonant transport display distinctive features. For resonant transport, the excitation can be transferred from the source to drain, while the ring track is moderately occupied. Far from resonance, the excitation oscillates between source and drain, minimmally populating of the ring on the the observed time scales. The nature of the oscillation depends on the characteristics of the ring.
Clearly, the far-off-resonance dynamics is  nearly independent on the detuning (indicating that the detuned sites are bypassed by the long-range hopping). By increasing $\alpha$, we observe a non trivial effect of detunings which is first localized around the resonant transport and then extended to the non-resonant dynamics. 

By studying the phase difference between upper and lower arms of the ring, we can conclude that the detuning can affect the phase of excitations (similarly to the effect of barriers in matter-wave propagation). The corresponding phase difference is observed to oscillate, even  for small population oscillations. Phase difference and population oscillation frequencies are related. 

Besides its own interest, we think our work is relevant to conceive devices of practical value employing Rydberg atoms or trapped ions.  
Finally, we point out that our logic  can be feasibly extended to systems with different geometries, also in the presence of noise.

\begin{acknowledgements}
{\it Acknowledgments.} We thank Giampiero Marchegiani and Tobias Haug for useful discussions.  The Julian Schwinger Foundation grant JSF-18-12-0011 is acknowledged. OM also acknowledges support by the H2020 ITN ``MOQS" (grant agreement number 955479) and MUR (Ministero dell’Università e della Ricerca) through the PNRR MUR project PE0000023-NQSTI. Numerical computations have been performed using the Julia package \texttt{QuantumOptics.jl}~\cite{kramer2018quantumoptics}.
\end{acknowledgements}

\appendix
\section{Distance and weak coupling limit}
\label{App:Dist}
\begin{figure*}[!t]
\centering
\includegraphics[width=0.8\textwidth]{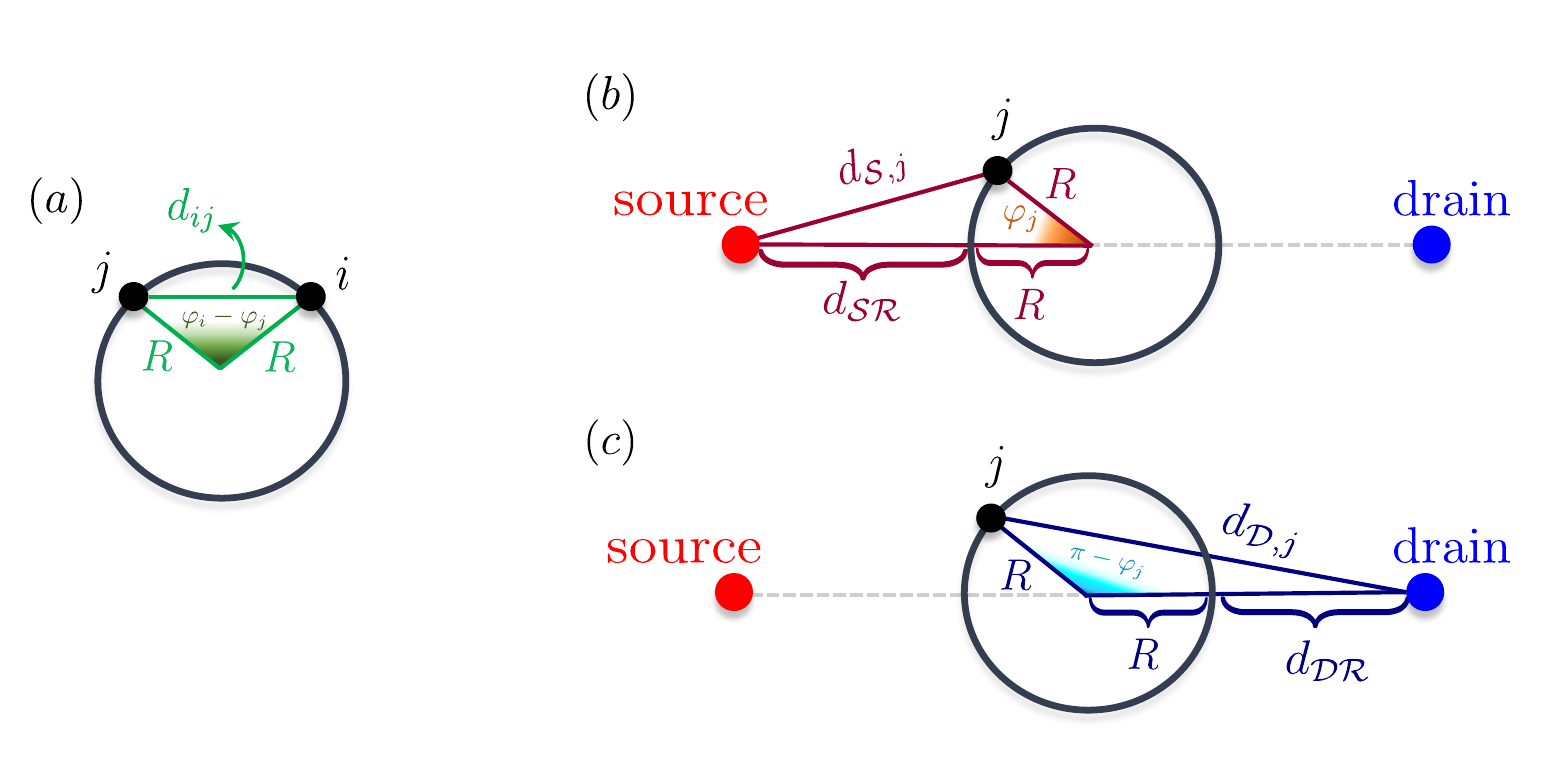}
\caption{Pictorial representation of distances in the system considered. Panel $(a)$: intra-ring distance between two sites located at positions $i$ and $j$. Panel $(b)$: source-ring distance. Panel $(c)$: drain-ring distance.}
\label{fig:distances}
\end{figure*}
Here we provide a proper definition of the intra-ring and ring-leads distances, then  we specify what we mean for weak ring-leads coupling. To define the distances, we first introduce an appropriate labeling of the atoms in the ring: supposing to work with an even number of sites in the ring, we label as $N_r$ the site closest to the source and $N_r/2$ the site closest to the drain. We identify with $\varphi_j=2\pi j / N_r$ the angle associated to site $j$ of the ring. Given this labeling, the source-ring distance is
\begin{subequations}
\begin{equation}
d_{\mathcal{S},j}=\sqrt{(d_{\mathcal{S},\mathcal{R}}+R)^2 + R^2 - 2(d_{\mathcal{S},\mathcal{R}}+R)R\cos\varphi_j},
\end{equation}
where $R$ is the radius of the ring and $d_{\mathcal{S},\mathcal{R}}$ is the distance between the source and the site $N_r$. Analogously, the drain-ring distance is
\begin{equation}
d_{\mathcal{D},j}=\sqrt{(d_{\mathcal{D},\mathcal{R}}+R)^2 + R^2 - 2(d_{\mathcal{D},\mathcal{R}}+R)R\cos(\pi - \varphi_j)},
\end{equation}
$d_{\mathcal{D},\mathcal{R}}$ being the distance between the drain and the site $N_r/2$. In both cases $j=1,\ldots, N_r$. The intra-ring distance is
\begin{equation}
    d_{i,j}=2R\sin\left( \dfrac{|\varphi_i - \varphi_j|}{2} \right)=2R\sin\left( \dfrac{\pi|i - j|}{N_r} \right)
\end{equation}
\end{subequations}
where $i,j=1,\ldots, N_r$. Finally, the source-drain distance is $d_{\mathcal{S},\mathcal{D}}=d_{\mathcal{S},\mathcal{R}}+2R+d_{\mathcal{D},\mathcal{R}}$. A sketch of the distances is reported in Fig.~\ref{fig:distances}.

Given a proper definition of the intra-ring and leads-ring distances, we introduce the intra-ring nearest-neighbor ($nn$) hopping
\begin{equation}
    J_{nn}=\dfrac{g}{d_{nn}^\alpha},
    \quad \mbox{with } \;\; d_{nn}=2R\sin\left(\dfrac{\pi}{N_r}\right),
\end{equation}
and the leads-ring nearest-neighbor hopping
\begin{equation}
    K_{nn}=\dfrac{g}{d_{\mathcal{S},\mathcal{R}}^{\alpha}}=\dfrac{g}{d_{\mathcal{D},\mathcal{R}}^{\alpha}},
\end{equation}
where we are assuming $d_{\mathcal{S},\mathcal{R}}=d_{\mathcal{D},\mathcal{R}}$. In the main text we work in the weak leads-ring coupling limit $K_{nn} \ll J_{nn}$, meaning $d_{\mathcal{S},\mathcal{R}}^{\alpha}\gg d_{nn}^{\alpha}$. Thus, if we consider $K_{nn}=J_{nn}/M$, $M$ being an integer bigger than one, the distances will be related by $d_{\mathcal{S},\mathcal{R}}=M^{1/\alpha}d_{nn}$.

\section{Ring spectral analysis}
\label{app:Ring_shift}

In the weak coupling regime, the ring and leads modes can be treated separately, their relation is crucial to understand the nature of the transport. The initial state of our protocol is a zero energy state, thus, the presence of zero energy modes in the ring is responsible for the population of the latter during the dynamics. In absence of zero energy modes in the ring, the latter will not be populated and the dynamics will result in coherent oscillations between source and drain. In this section, we will study the effect of localized detunings on the spectrum of an XY model in a ring. We first use a perturbative approach to study the effect of the detunings on the spectrum of the nearest-neighbor model; it is instructive to understand the possible presence of resonances between ring and drain. Then, we numerically analyze the modification of the ring Hamiltonian spectrum due to the long-range hopping paying particular attention to the changes in terms of resonances with zero energy modes. We work in the one excitation sector of the Hamiltonian.

\subsection{Nearest-neighbor XY energy shift due to detuning}
Let us consider the generic Hamiltonian 
\begin{equation}
    \hat{\mathcal{H}} = \hat{\mathcal{H}}_0 + \lambda \hat{\mathcal{H}'},
\end{equation}
where $\hat{\mathcal{H}}_0$ is the unperturbed Hamiltonian, while $\hat{\mathcal{H}'}$ is a perturbation whose intensity $\lambda$ is supposed to be small with respect to the unperturbed eigenvalues. We consider a Hamiltonian that possesses at least a couple of degenerate states $\ket{E_a^0}$ and $\ket{E_b^0}$ such that
\begin{equation}
\hat{\mathcal{H}}_0\ket{E_{a,b}^0}=E^{(0)}\ket{E_{a,b}^0}
\end{equation}
and work in the subspace spanned by these two states. Let us introduce the matrix element
\begin{equation}
W_{ij}=\braket{E_i^0|\hat{\mathcal{H}'}|E_j^0}.
\end{equation}
The first order correction to the unperturbed eigenvalue is~\cite{griffiths2018introduction}
\begin{align}
    &E^{(1)}_{\pm} = \frac{1}{2} \Big[ W_{aa}+W_{bb}\pm \sqrt{(W_{aa}-W_{bb})^2+4|W_{ab}|^2} \Big], \nonumber \\
    &E_{\pm} = E^{(0)} + \lambda E^{(1)}_{\pm}.
    \label{eq:correction_1st}
\end{align}
Now, we consider a XY model on a ring composed of $N_r$ sites perturbed by a localized detuning term:
\begin{equation}
    \hat{\mathcal{H}} = J\sum_{j=1}^{N_r}\left(\hat{\sigma}_j^+ \hat{\sigma}_{j+1}^- + \operatorname{H.c.}\right) + \Delta\sum_{j=1}^{N_r}\hat{n}_j\delta_{j,j_0} = \hat{\mathcal{H}}_0 + \Delta \hat{\mathcal{H}'}
\end{equation}
where here $\Delta$ is the perturbative parameter and we work in the one excitation sector. The unperturbed Hamiltonian can be diagonalized by transforming the spins in fermions via a Jordan Wigner transformation~\cite{jordan1993paulische, de2009x}
\begin{equation}
\label{eq:c-fermions}
    \hat{c}_j = \bigg( \prod_{\ell<j}\hat{\sigma}_{\ell}^{z} \bigg) \hat{\sigma}_j^-,
\end{equation}
such that the $\hat c^{(\dagger)}_j$ operators obey the usual anticommutation relations for fermions.
The resulting unperturbed model is a fermionic non-diagonal Hamiltonian. In the one excitation sector, the fermionic operators are rotated through the transformation
\begin{equation}
    \hat{c}_j^{\dagger} = \dfrac{1}{\sqrt{N_r}}\sum_{n=1}^{N_r}e^{2\pi i j n / N_r} \hat{d}_n^{\dagger}
\end{equation}
and the diagonal unperturbed Hamiltonian is obtained~\cite{de2009x,zawadzki2017symmetries}
\begin{equation}
    \hat{\mathcal{H}}_0 =  \sum_{n=1}^{N_r}E_n^{(0)}\hat{d}_n^{\dagger}\hat{d}_n,
    \quad
    E_n^{(0)} = 2J \cos\left(\dfrac{2\pi n}{N_r} \right).
\end{equation}
Defining the momentum $k=2\pi n / N_r$, we can observe that the eigenvalues of the Hamiltonian are all doubly degenerate, except for the ground state and the maximum excited state. We can easily observe that the couples of degenerate eigenstates are of the form $\ket{k}$ and $\ket{2\pi - k}$ (or alternatively $\ket{n}$ and $\ket{N_r - n}$) which refers to states with momenta $k$ and $2\pi-k$; thus, we can restrict our analysis in the subspace spanned by these two states. Here, the perturbation matrix element is
\begin{equation}
    W_{k,k'}=\braket{k|\hat{n}_{j_0}|k'}.
\end{equation}
To perform the computation it is convenient to transform the number operator in terms of Jordan Wigner fermions in the momentum space. Done that, the first matrix element to which we are interested is 
\begin{eqnarray}
    W_{k,k} & = & \dfrac{1}{N_r}\sum_{m,m'=1}^{N_r}e^{-2\pi i m j_0/ N_r}e^{2\pi i m' j_0/ N_r}\braket{n|\hat{d}_m^{\dagger}\hat{d}_{m'}|n} \nonumber\\
    & = & \dfrac{1}{N_r}\sum_{m,m'=1}^{N_r}e^{2\pi i (m'-m) j_0/ N_r}\delta_{n,m}\delta_{m',n}=\dfrac{1}{N_r}\nonumber\\
    & = & W_{2\pi - k,2\pi - k}.
\end{eqnarray}
At the same time, the off-diagonal term will be
\begin{eqnarray}
    W_{k,2\pi-k} & = & \dfrac{1}{N_r}\sum_{m,m'=1}^{N_r}e^{2\pi i (m'-m) j_0/ N_r}\delta_{n,m}\delta_{m',N_r-n} \nonumber\\
    & = & \dfrac{e^{2ikj_0}}{N_r}=W^{*}_{2\pi-k,k}.
\end{eqnarray}
Given the perturbation matrix elements, we can compute the first order energy corrections using eq.\eqref{eq:correction_1st}
\begin{equation}
        E_{\pm} = E^{(0)} + \Delta \dfrac{2\pm 2}{2N_r}.
\end{equation}
Thus, one of the two degenerate states remains unchanged and the other one is shifted by a factor $2\Delta/N_r$. The non degenerate eigenvalues $E_{GS}$ and $E_{max}$ are shifted by a common factor $W_{k_{GS},k_{GS}}=\braket{k_{GS}|\hat{n}_{j_0}|k_{GS}}=W_{k_{max},k_{max}}=\braket{k_{max}|\hat{n}_{j_0}|k_{max}}=\Delta/N_r$, where $k_{GS}$ and $k_{max}$ are the momenta associated respectively to the ground and the maximum energy state. 

\begin{figure}[!t]
\centering
\includegraphics[width=1\columnwidth]{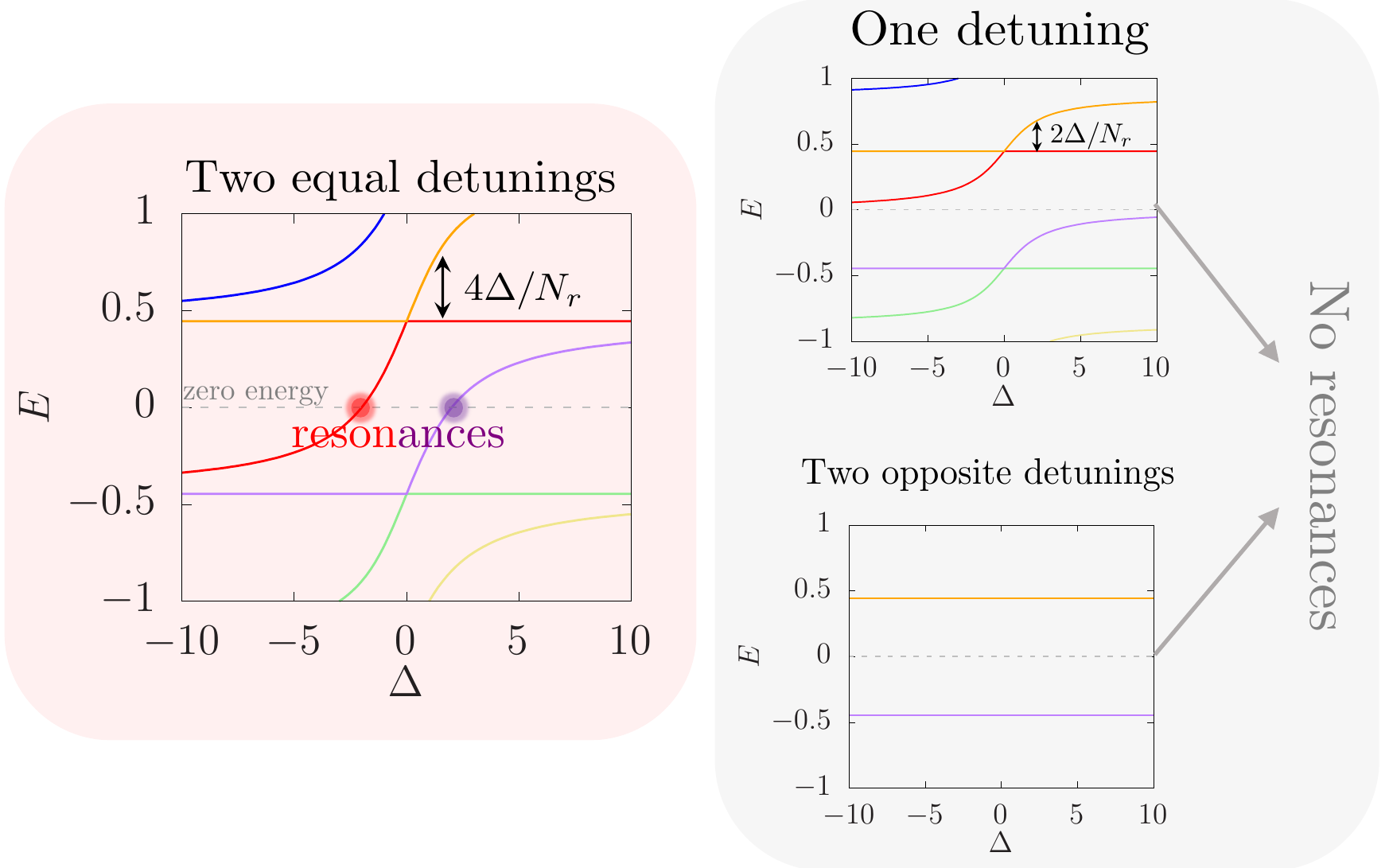}
\caption{Numerical results for the spectrum of the nearest-neighbor XY model in presence of localized barriers. Only energies with energy near to zero are reported. The number of sites is $N_r=14$ and $J=1$.}
\label{fig:shiftNN}
\end{figure}

In the case of two detunings with the same sign, the perturbation Hamiltonian is
\begin{equation}
\hat{\mathcal{H}}'=\hat{n}_{j_A}+\hat{n}_{j_B};
\end{equation}
the perturbation matrix elements are
\begin{subequations}
\begin{align}
    W_{k,k}  = & \dfrac{2}{N_r}=W_{2\pi-k,2\pi-k}, \\
    W_{k,2\pi-k}  = & \dfrac{1}{N_r}\left( e^{2ikj_A}+e^{2ikj_B}\right)=W^{*}_{2\pi-k,k}.
\end{align}
\end{subequations}
Given the matrix elements, we can compute the energy correction for degenerate states via
\begin{equation}
\eqref{eq:correction_1st} 
        E_{\pm} = E^{(0)} + \frac{\Delta}{N_r} \Big\{ 2\pm \sqrt{2+2\cos \big[ 2k(j_A - j_B) \big]} \Big\}.
\end{equation}
If we set the barriers in opposite sides of the ring, i.e. $j_A=\ceil{N_r/4}$ and $j_B=\ceil{3N_r/4}$, in such a way that $|j_A-j_B|=N_r/2$, we obtain a shift of a factor $0$ or $4\Delta/N_r$ for each state. On the other hand, the ground and the maximum energy state are both shifted by a factor $2\Delta/N_r$.

Finally, it is straightforward to observe that the first order correction to the energy states in presence of two barriers with opposite sign gives zero contribution at the first perturbative order. To do that, it is sufficient to perform the same calculation done for two equal detunings with $\hat{\mathcal{H}}'=\hat{n}_{j_A}-\hat{n}_{j_B}$. If $|j_A-j_B|=N_r/2$, the first order energy correction is zero for each state of the spectrum. 

Figure~\ref{fig:shiftNN} reports the numerical results for the energy shifts in presence of localized detunings. Through the numerical computation we easily understand what happens beyond the first perturbative order. In particular, both for single and double equal detunings, the first order shifts in $\Delta$ are corrected for big values of the latter tending to flatten. The main difference between the two cases is the slope of the shift at small values of the detuning. The one of the single barrier case is smaller than the one of the double barrier case. Thus, while in the latter it is sufficient to permit to two levels to cross the zero energy mode, in the single barrier case the linear shift it is not sufficient. It is important to observe that we do not observe resonances in our $\Delta$ range, it is possible that they are present for bigger values of the detuning. However, in the case of two opposite detunings, the degenerate states reported are not shifted even for more than first order contributions; then, we do not expect resonances, the dynamics is characterized by coherent oscillations between source and drain.

\begin{figure*}[!t]
\centering
\includegraphics[width=0.85\textwidth]{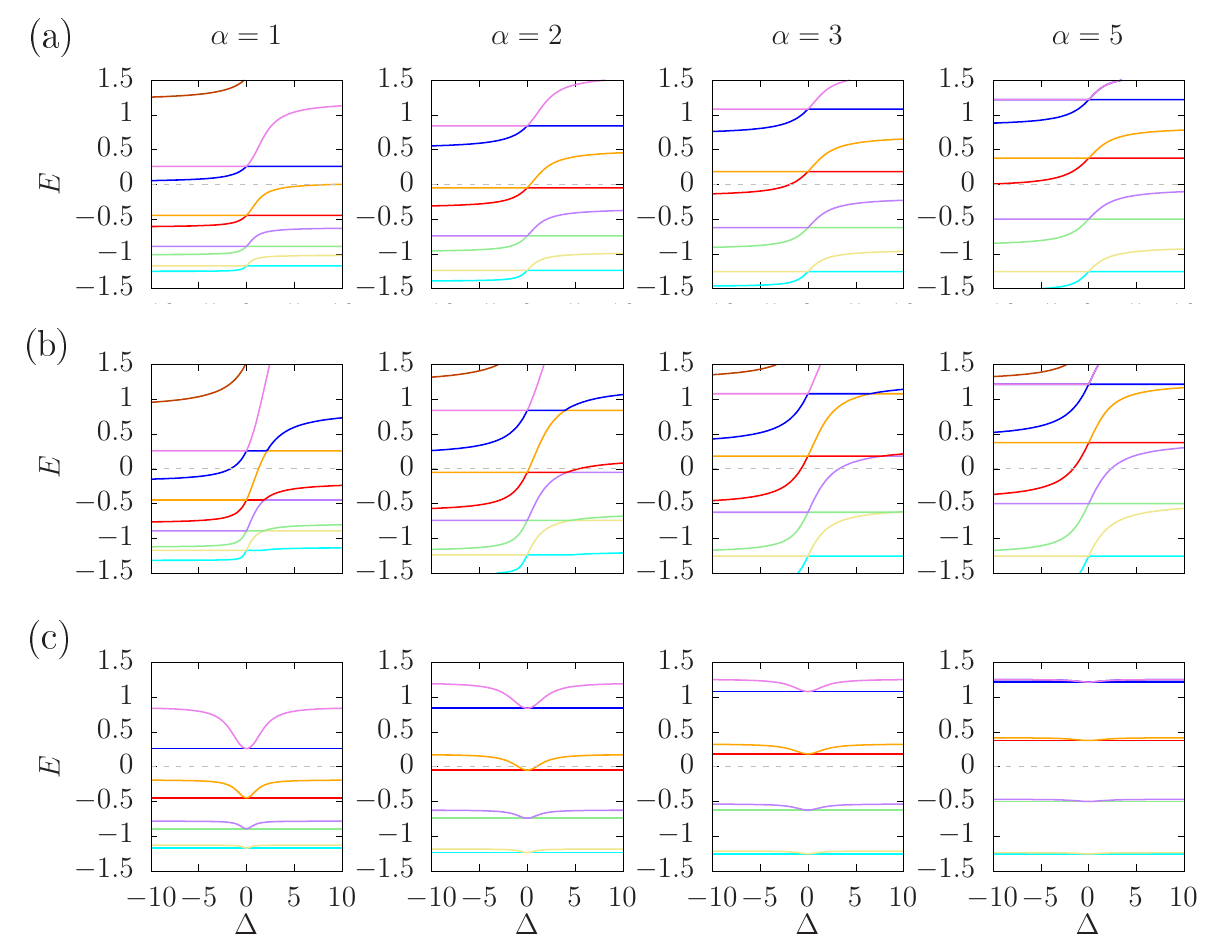}
\caption{Numerical results for the spectrum of the long-range XY model in the ring in presence of localized detunings \eqref{eq:Ring}. Only close to zero energies are reported. The plot is organized in three panels: panel (a) reports results in presence of a single detuning, panel (b) reports results in presence of two equal detunings and panel (c) reports results in presence of two opposite detunings. For each panels we show results for $\alpha=1,2,3,5$. The number of sites is $N_r=14$ and $J_{nn}=1$.}
\label{fig:shiftLR}
\end{figure*}

\subsection{Long-range XY energy shift due to detuning}

For a nearest-neighbor XY model, the spectrum is symmetric with respect to the zero energy states, meaning that the energies are organized in pairs $\pm E_n$. We also showed how from the energy shifts we expect that the resonances are symmetric for $\Delta\rightarrow -\Delta$. However, the dynamics reported in the main text shows that for long-range hopping this is not true. The ring population does not follow a symmetric behavior with respect to $\Delta=0$. Therefore, a detailed analysis of the energy shift in the spectrum is important to understand the asymmetry in the dynamics. The model with which we work here is \eqref{eq:Ring}.

Figure~\ref{fig:shiftLR} reports the energy spectrum in all the three paradigmatic detuning configurations considered so far. Before analyzing the effect of the detuning, it is necessary to pay attention on the role of the inverse hopping range $\alpha$ for $\Delta=0$. We can immediately observe that for each value of $\alpha$ the degeneracies persist; on the other hand, a substancial shift of the levels is observed. Passing from $\alpha=5$ to $\alpha=1$ the levels lose their symmetry with respect to $E=0$ and tend to crush to a negative energy value. This means that the first $E>0$ and the first $E<0$ states does not possess anymore symmetric energy with respect to zero. When the detuning is switched on, the energy levels are shifted and cross the zero energy mode.

To be more quantitative, let us first consider the case of the single barrier reported in Fig.~\ref{fig:shiftLR}(a). Here, the levels that cause resonances are the orange and red. Increasing the hopping range, the red and the orange levels are shifted downwards; during the shift, one of the two levels crosses the zero energy mode giving rise to one resonance at negative $\Delta$. The purple level in the nearest-neighbor limit is not sufficiently shifted to give rise to a positive resonance at positive $\Delta$ in the interval considered so far. Then, for one detuning, we have only one value or narrow range of $\Delta$ for which the ring becomes populated during the dynamics.

\begin{figure}[!t]
\centering
\includegraphics[width=0.85\columnwidth]{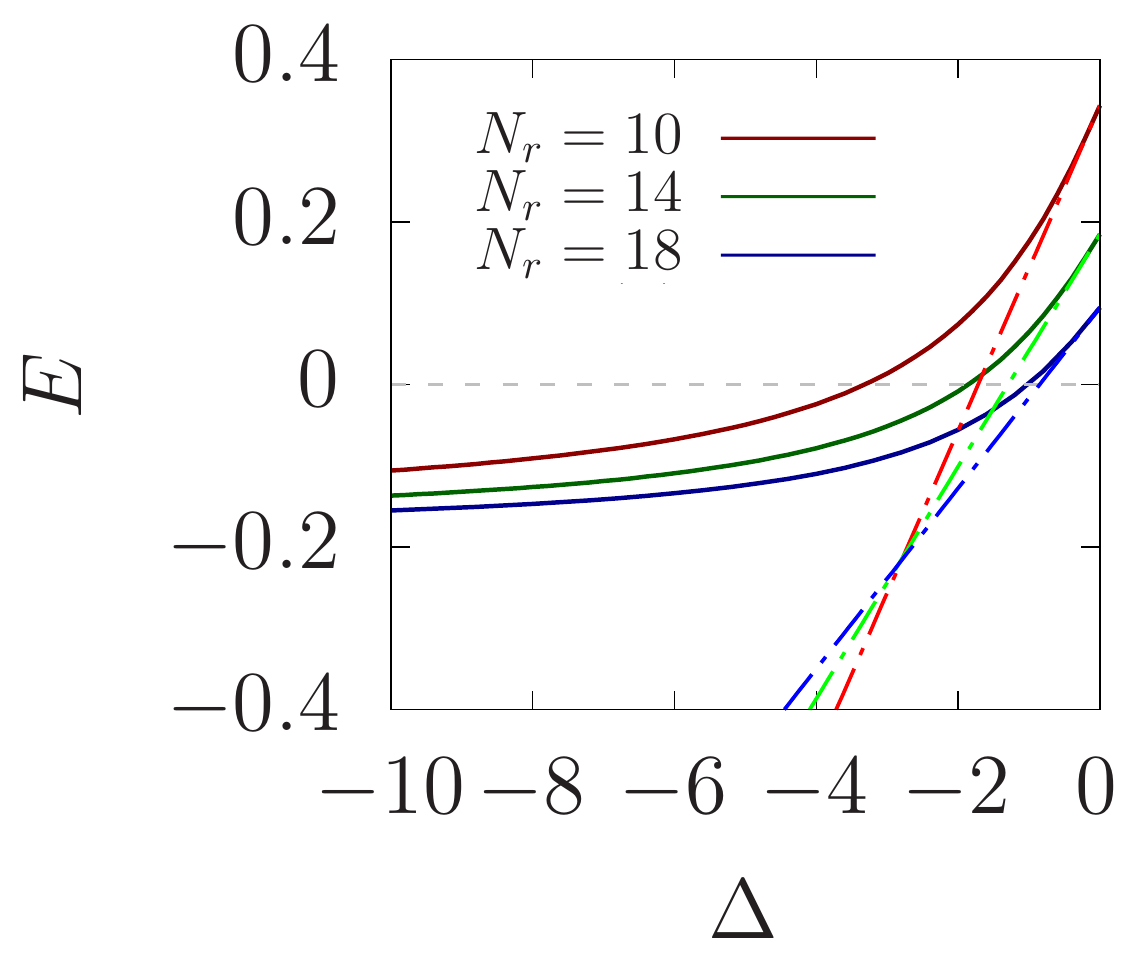}
\caption{Crossing energy states in presence of a single barrier located at $j_0=\ceil{N_r/4}$ for three different values of the number of atoms in the ring $N_r$. The solid lines refers to numerical values of the energy for $N_r=10$ (dark red), $N_r=14$ (dark green) and $N_r=18$ (dark blue). The dashed blue, green and blue lines refers to first perturbative order corrections $E^{(0)}+2\Delta/N_r$ for respectively $N_r=10,14,18$. In all the three cases $\alpha=3$ and the parameters are set in such a way that $J_{nn}=1$.}
\label{fig:shiftLR_N}
\end{figure}

Figure~\ref{fig:shiftLR}(b) reports the results for two equal detunings: for each value of $\alpha$ there are at least two values of $\Delta$ for which the zero energy mode is crossed, as in the nearest-neighbor case. In presence of two opposite detunings [Fig.~\ref{fig:shiftLR}(c)], the long-range hopping is responsible for a shift of the energy levels that is quadratic in the detuning $\Delta$. As in the nearest-neighbor limit, also in presence of long-range the first order correction is zero, thus the shift scales as $\Delta^2$ with a coefficient that increases as $\alpha$ decreases. For this reason, it is possible to have symmetric resonances with respect to $\Delta=0$. In particular, in the plots reported, resonances appear at $\alpha=2$.

It is safe to assume that the presence of the resonances strongly depends also on the number of atoms in the ring. The presence of size dependent resonance location can be immediately deduced from the perturbative results obtained in the previous section. Indeed, the unperturbed energy is $E^{(0)}_k = 2J \cos k$ where $k\sim 1/N_r$ and the first order shifts are directly proportional to $\Delta/N_r$. In presence of long-range hoppings we do not have access to exact spectrum of the Hamiltonian; however, at least for finite value of $N_r$ we can analyze how the resonance position moves increasing the size. To be more precise, we consider the case of single barrier and plot the crossing states in function of $\Delta$. Here, for crossing states we mean the states that cross the zero energy level in the interval $\Delta\in [-10,0]$. 

Figure~\ref{fig:shiftLR_N} reports them and the first order linear corrections $E^{(0)}+2\Delta/N_r$, where we extracted $E^{(0)}$ from numerical data and assumed that the the first order perturbative shift  is unchanged with respect to the nearest-neighbor case. We immediately observe that for small values of $\Delta$ the linear shift keeps working well; indeed, increasing the size of the system, the slope of the shift is smaller. Moreover, the energy $E^{(0)}$ is smaller increasing the size and this makes the resonance position located at smaller values of $|\Delta|$ for bigger values of the size. Finally, we observe that already for $N_r=18$ the position of the resonance can be qualitatively estimated from the linear shift as $\Delta_{res} \approx -E^{(0)}N_r/2$, since it is located at a relatively small value of the detuning. 

\section{Dynamics in absence of detunings}
\label{app:NoBarr}
\begin{figure}[!h]
\centering
\includegraphics[width=\columnwidth]{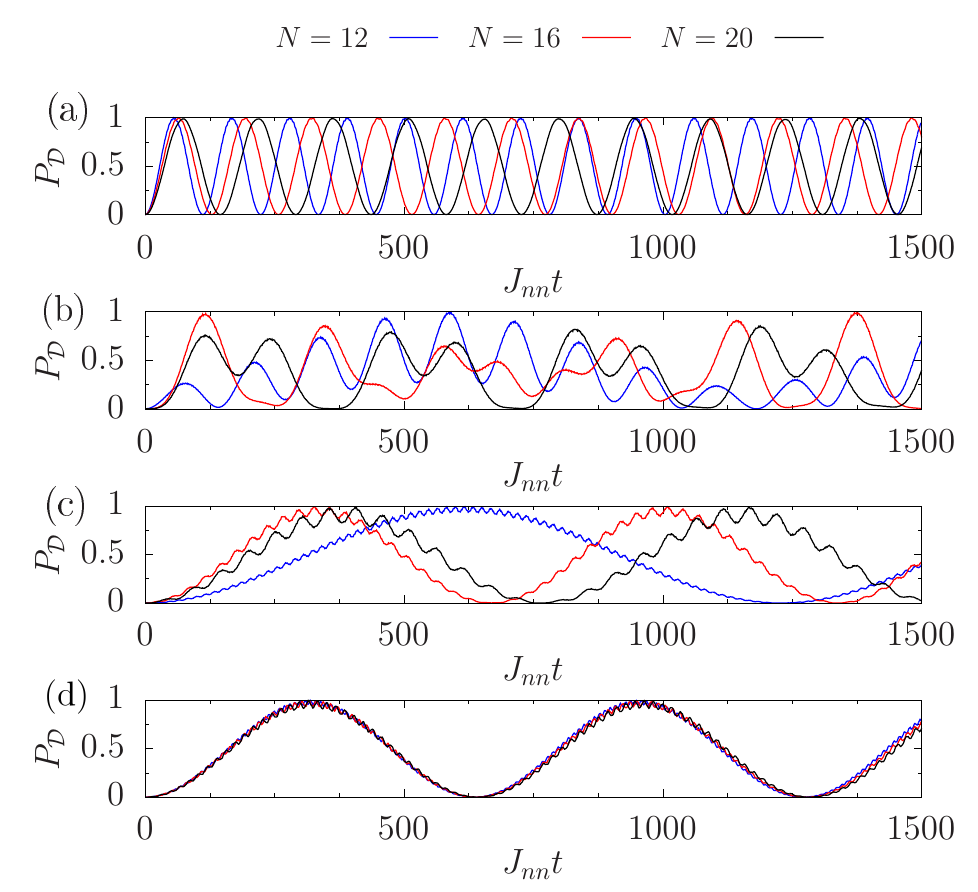}
\caption{Dynamics of the drain population in absence of detunings in the ring. Results are reported for three different values of the size of the system: $N=12$ (blue), $N=16$ (red) and $N=20$ (black). The four panels refers to different values of the inverse hopping range: $\alpha=1$ in panel (a), $\alpha=3$ in panel (b), nearest-neighbor in panel (d). Panel (b) shows the dynamics for the values of the inverse hopping range for which the ring presents zero energy levels in absence of detunings. These values are $\alpha_{res}=1.731$ ($N=12$), $\alpha_{res}=2.168$ ($N=16$), $\alpha_{res}=2.508$ ($N=20$). Leads and ring are weakly coupled $K_{nn}/J_{nn}=0.1$.}
\label{fig:PvsD_zeroBar}
\end{figure}
Here we analyze the dynamics of the system in absence of any type of detuning. We expect to find a non trivial dynamics also in absence of it; due to the geometry of the system, the excitation that travels along the system is subject to many scattering and splitting events. To have an idea of the complexity of the dynamics, let us consider an excitation subject only to nearest-neighbor hopping that travels in the system starting from the source. The excitation moves from the source to the nearest site on the ring. The coupling between source and ring is smaller compared to the intra-ring one, thus, once a small excitation fraction reaches the ring, it is quickly split in the two arms of the ring. The splitted excitation fractions travel in the two arms of the ring and meet at the end of them, before the drain. Here, they scatter, a part of the excitation hops to the drain, the rest continues its dynamics in the arms of the ring. During a long time evolution a lot of scattering events occur and the dynamics will result complex and not intuitively easily to describe.

Figure~\ref{fig:PvsD_zeroBar} reports the population dynamics of the drain for different values of the inverse hopping range $\alpha$. we consider $\alpha=1$, $\alpha=3$, nearest-neighbor limit and the particular value of $\alpha$ for which, in absence of detuning, the ring presents zero energy levels resonant with the initial state. For instance, from Fig.\ref{fig:shiftLR} we can observe that for $N=16$ the particular $\alpha_{res}$ value is close to $2$. Indeed, we find $\alpha_{res}=2.168$; the other $\alpha_{res}$ values are reported in the Fig.\ref{fig:PvsD_zeroBar} caption. We observe how the drain population oscillates quite regularly except in the $\alpha_{res}$ case; the ring becomes is well populated and the oscillations are extremely irregular. For the other cases the ring is weakly populated, see Figs.\ref{fig:3dplots_singbarr},\ref{fig:3dplots_2eqbarr},\ref{fig:3dplots_2opbarr}, thus it is important to follow the behavior of the drain density. 

The latter is characterized by small frequency oscillations with big amplitude modulated by fast oscillations with small amplitude. In the nearest-neighbor limit, the two frequencies can be evaluated analytically~\cite{haug2019aharonov}
\begin{widetext}
\begin{equation}
    \dfrac{\omega_{\pm}}{J_{nn}}=\pm\dfrac{N_r}{\delta}\dfrac{8+\tilde{K}_{nn}^2 N_r - 2 \tilde{K}_{nn}^2 \delta }{4\tilde{K}_{nn}^2\delta^2 - N_r(8+\tilde{K}_{nn}^2 N_r)} + \sqrt{   
    \left(\dfrac{N_r}{\delta}\dfrac{8+\tilde{K}_{nn}^2 N_r - 2 \tilde{K}_{nn}^2 \delta }{4\tilde{K}_{nn}^2\delta^2 - N_r(8+\tilde{K}_{nn}^2 N_r)}   \right)^2
    + \dfrac{N_r}{\delta}\dfrac{8\tilde{K}_{nn}^2}{8N_r + \tilde{K}_{nn}^2(N_r^2-4\delta^2)} 
          }
\end{equation}
\end{widetext}
where $\tilde{K}_{nn}=K_{nn}/J_{nn}$ and $\delta=\csc(\pi/N_r)$. From the numerical computation we can observe that the frequencies are weakly dependent on the size of the system. The non trivial form of the frequencies confirms how the dynamics is complex.

Surprisingly, for $\alpha=3$ the dynamics changes a lot with respect to the nearest-neighbor limit. One of the crucial differences, is that here the leads are strongly connected with a plethora of sites near to the $N_r$ and $N_r/2$, not just with one site. The first counter-intuitive result is the slow down of the dynamics with respect to the nearest-neighbor case. The presence of hoppings does not cause a faster transport of the excitation from one site to the other. Secondly, the transport is slower for $N=12$, while $N=16$ is the faster case. Thus, a smaller size does not correspond to faster dynamics. Lastly, the case $\alpha=1$ is the most intuitive one. The hopping range is sufficiently large to avoid complex internal effects in the ring. The oscillations are faster with respect to all the other cases considered with smaller hopping range; moreover, increasing the system size it becomes slower.

\section{Information on the dynamics from the complete spectrum}
\label{app:allspectrum}

In~\ref{app:Ring_shift} we showed that important information on the dynamics can be extracted from the bare ring Hamiltonian, assuming that the ring and the leads are weakly coupled. Here, we focus our attention on the complete Hamiltonian, describing how the resonances manifest themselves on the state population and how the transport nature is modified and damaged from strong coupling and disorder.

\subsection{Full Hamiltonian eigenstate nature}
In the one excitation sector, a generic eigenstate of the Hamiltonian can be written in the computational basis as 
\begin{equation}
\ket{E_{n}}=c_{\mathcal{S}}^{(n)}\ket{\mathcal{S}}+\ket{\mathcal{R}}+c_{\mathcal{D}}^{(n)}\ket{\mathcal{D}},
\end{equation}
where $\ket{\mathcal{S}}=\ket{\uparrow}_{\mathcal{S}}\otimes\ket{\downarrow,...,\downarrow}_{\mathcal{R}}\otimes \ket{\downarrow}_{\mathcal{D}}$, $\ket{\mathcal{R}}=\sum_{j\in \mathcal{R}}c_{j}^{(n)}\ket{\downarrow}_{\mathcal{S}}\otimes\ket{\downarrow,...,\uparrow_{j},...,\downarrow}_{\mathcal{R}}\otimes \ket{\downarrow}_{\mathcal{D}}$ and $\ket{\mathcal{D}}=\ket{\downarrow}_{\mathcal{S}}\otimes\ket{\downarrow,...,\downarrow}_{\mathcal{R}}\otimes \ket{\uparrow}_{\mathcal{D}}$ are respectively source, ring, and drain states. In the absence of resonances, it is safe to assume that the leads and the ring degrees of freedom are separated. Thus, we group the states of the Hamiltonian in three classes: $\{ \ket{E_{n_{l}}}, \ket{E_{n_{r}}}, \ket{E_{n_{lr}}} \}$, where $\ket{E_{n_{l}}}$ are states in which all the excitations are in the leads, $\ket{E_{n_{r}}}=\ket{\mathcal{R}}$ are states in which only the ring is populated, and $\ket{E_{n_{lr}}}$ are states in which the population is distributed between ring and leads. The labels satisfy the relation $n_{l}+n_{r}+n_{lr}=N$. Since the initial state of our protocol is $\ket{\psi(0)}=\ket{\mathcal{S}}$ and $\braket{\mathcal{R}|\mathcal{S}}=0$, the states $\ket{E_{n_{r}}}$ will not contribute to the dynamics; thus, the generic time evolved state can be written as
\begin{eqnarray}
    \ket{\psi(t)} &=& \sum_{n_{l}} e^{-iE_{n_{l}}t}\braket{E_{n_{l}}|\mathcal{S}}\ket{E_{n_{l}}} \nonumber\\
    & + &\sum_{n_{lr}} e^{-iE_{n_{lr}}t}\braket{E_{n_{lr}}|\mathcal{S}}\ket{E_{n_{lr}}}.
\end{eqnarray}
Moreover, if the Hamiltonian spectrum does not contain uniform states $\ket{E_{n_{lr}}}$, the dynamics is dominated by the leads states $\ket{E_{n_{l}}}=c_{\mathcal{S}}^{(n_{l})}\ket{\mathcal{S}}+c_{\mathcal{D}}^{(n_{l})}\ket{\mathcal{D}}$ and will result in oscillations between source and drain of the form
\begin{equation}
        \ket{\psi(t)}=\sum_{n_{l}} e^{-iE_{n_{l}}t}\left(c_{\mathcal{S}}^{(n_{l})}\right)^{*}\left[c_{\mathcal{S}}^{(n_{l})}\ket{\mathcal{S}}+c_{\mathcal{D}}^{(n_{l})}\ket{\mathcal{D}}\right].
        \label{eq:lead_state}
\end{equation}
The value of the lead state energies and their coefficients is fundamental to have information on the source-drain oscillation frequency and amplitude. On the other hand, the ring becomes populated only when the Hamiltonian has $\ket{E_{n_{lr}}}$ states. Thus, we expect to observe states $\ket{E_{n_{lr}}}$ in the spectrum in correspondence of the resonances.

Here we focus our attention on the eigenstate population $P^{(n)}_j=\braket{E_n|\hat{n}_j|E_n}$, we compute it for source, ring and drain in the whole spectrum. 
\begin{figure*}[!t]
\centering
\includegraphics[width=0.85\textwidth]{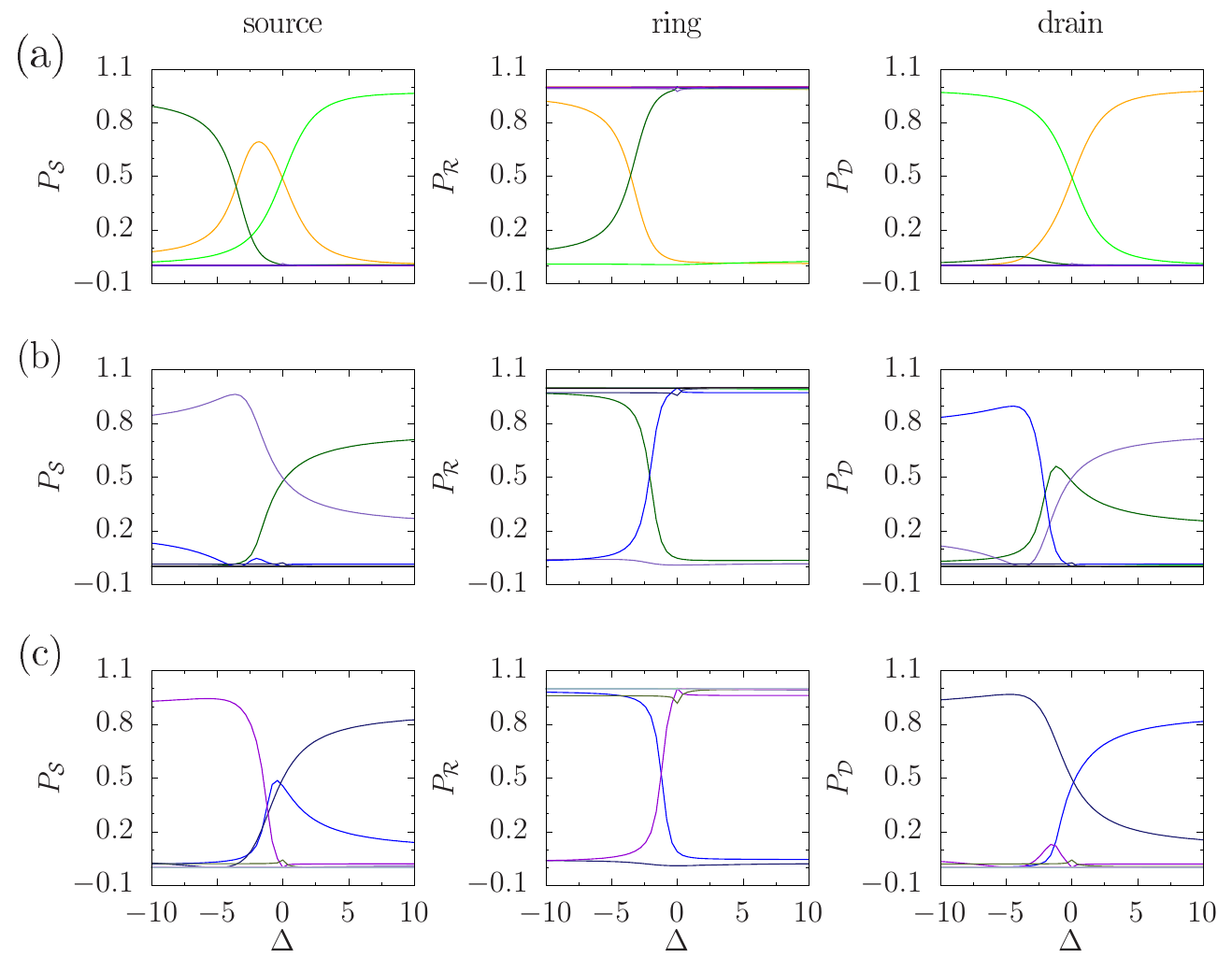}
\caption{Source, ring and drain population from left to right in the spectrum for three different values of the size $N=N_r+2$: $N=12$ (panel (a)), $N=16$ (panel (b)) and $N=20$ (panel (c)). Most of the states are approximately ring states, meaning that their $P_{\mathcal{R}}^{(n)}\approx 1$. For each $N$ only three states are not ring states: for $N=12$ the $5$-th, $6$-th and $7$-th (orange, green and dark green) excited states, for $N=16$ the $7$-th, $8$-th and $9$-th (dark green, purple and blue) excited states, for $N=20$ the $9$-th, $10$-th and $11$-th (blue, dark blue and violet) excited states. In all the three plots the parameters are set in such a way that $J_{nn}=1$ and $K_{nn}=0.1$, $\alpha=3$. Only one barrier is taken into account.}
\label{fig:PvsD}
\end{figure*}
Figure~\ref{fig:PvsD} reports the behavior of the eigenstate populations in leads and ring for three different values of the size of the system, the inverse hopping range is $\alpha=3$ and only one localized barrier at $j_0=\ceil{N_r/4}$ is taken into account. We immediately observe that the majority of the states of the system are ring states for which $P_{\mathcal{R}}^{(n)}\approx 1$ and $P_{\mathcal{S},\mathcal{D}}^{(n)}\approx 0$, they do not contribute to the dynamics if the system is initialized in $\ket{\mathcal{S}}$. Comparing between Fig.~\ref{fig:shiftLR_N} and Fig.~\ref{fig:PvsD}, we observe that in correspondence of the resonance the ring population is $1/2$ for two states of the spectrum, the excitation is equally distributed between ring and leads and this will result in a filling of the ring during the dynamics. Far from resonance, the Hamiltonian states are approximately ring or leads states with $P_{\mathcal{R}}^{(n)}\approx 0,1$; thus, the dynamics will result in coherent oscillations between source and drain.

Let us observe that leads states possess $P_{\mathcal{L}}^{(n)}=P_{\mathcal{S}}^{(n)}+P_{\mathcal{D}}^{(n)}=1$, therefore, the nature of the dynamics strongly depends on the source and drain population of the states. In particular, for $\Delta < \Delta_{res}$ there are source localized states, meaning states in which the source population is dominant an all the others. In absence of uniform states, the dynamics follows Eq.~\eqref{eq:lead_state}, that results in $\ket{\psi(t)}\approx e^{i\phi}\ket{\mathcal{S}}$ if there is a dominant localized source state. Thus, for $\Delta < \Delta_{res}$ the excitation remains localized in the source or at most a small fraction of it moves to the drain. The value of the population shows an irregular dependence on the size, the position of the resonance is size dependent and various finite size effects comes into play. Far from resonance, the size dependence is not regular, the source and drain populations follows different behaviors without a well defined dependence on the size. Thus, for each value of the size there is a resonance, but the dynamics far from the resonance strongly depends on $N$.

\subsection{Eigenstate leads population for different leads-ring coupling}
\label{app:strong_spectrum}
\begin{figure*}[!t]
\centering
\includegraphics[width=0.85\textwidth]{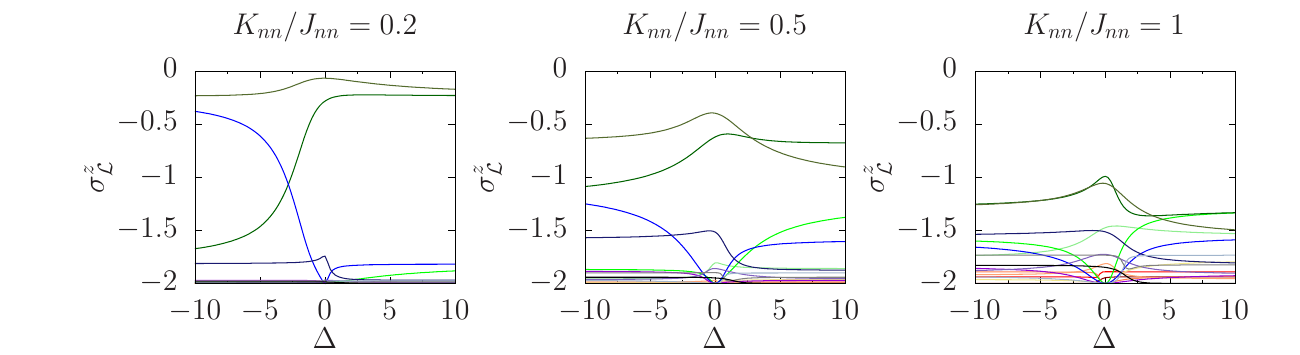}
\caption{Leads magnetization in the spectrum in function of the barrier height $\Delta$ for different leads-ring couplings $K_{nn}/J_{nn}$. $J_{nn}=1$ is fixed and $K_{nn}$ is changed. $N_r=14$ and $\alpha=3$ are considered. The ring contains a single barrier localized in $\ceil{N_r/4}$.}
\label{fig:diff_coupling}
\end{figure*}
In the previous subsection we analyzed the full spectrum populations in the weak coupling regime. Here, we study it far from the weak coupling regime. We study the system in terms of leads population, in particular we analyze the leads magnetization that is directly related to leads and ring populations
\begin{equation}
    \hat{\sigma}_{\mathcal{L}}^z = \hat{\sigma}_{\mathcal{S}}^z + \hat{\sigma}_{\mathcal{D}}^z = 2\big(\hat{n}_{\mathcal{L}} - \hat{1}\big) = -2\hat{n}_{\mathcal{R}}.
\end{equation}
This means that leads and ring states have respectively $\braket{\hat{\sigma}_{\mathcal{L}}^z}=0,-2$.

We always focused on the Rydberg case $\alpha=3$, Fig.~\ref{fig:diff_coupling} reports the leads magnetization for three different values of the leads-ring coupling. Increasing the coupling the population of the eigentstates changes drastically. For small coupling three states are $\ket{E_{n_{rl}}}$ states around the resonance point, they are the $7$-th, $8$-th and $9$-th excited states. The others are approximatively ring or leads states. Increasing $K_{nn}/J_{nn}$, both $\ket{E_{n_{r}}}$ and $\ket{E_{n_{l}}}$ states become $\ket{E_{n_{rl}}}$ states. The spectrum is not anymore separated in well defined classes, this will result in a dynamics that is not characterized by coherent source-drain oscillation, but the ring will be populated not only for specific $\Delta$ values. Thus, the transport will not be anymore controllable by tuning $\Delta$.

\subsection{Robustness against noise}
\begin{figure*}[!t]
\centering
\includegraphics[width=0.85\textwidth]{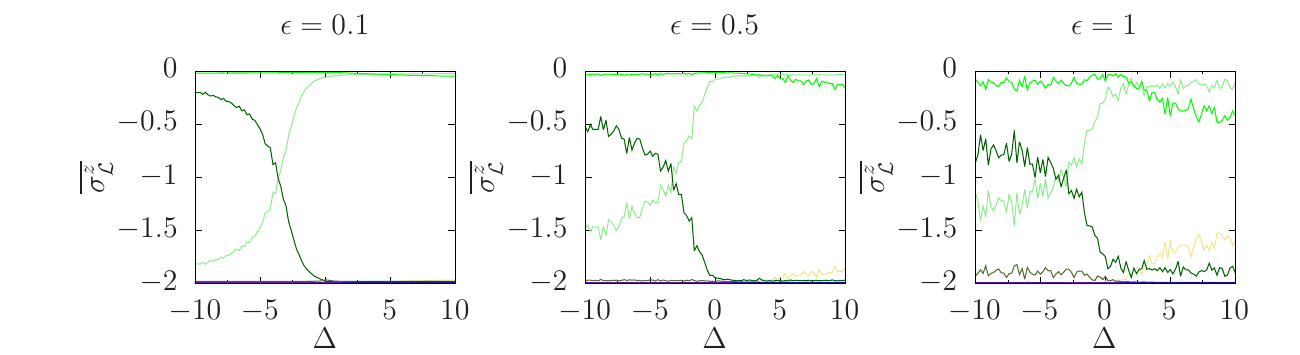}
\caption{Leads magnetization in the spectrum in function of the barrier height $\Delta$ for different values of the disorder $\epsilon$. $J_{nn}=1$ is fixed and $K_{nn}=0.1$ (weak coupling). $N_r=10$ and $\alpha=3$ are considered. The impurity Hamiltonian is \ref{eq:Ham_dis_1bar}. On the $y$ axis it is reported the expectation value \ref{eq:Magnetz} for each $n$, it is reported as $\overline{\sigma^z_{\mathcal{L}}}$. $N_{rea}=100$ disorder realizations are taken into account.}
\label{fig:noiseN12}
\end{figure*}
Here we study the robustness of the results under the presence of disorder. In particular, we work with the impurity Hamiltonian
\begin{equation}
        \hat{\mathcal{H}}_{imp}=\sum_{j\in \mathcal{R}}\Delta_j \hat{n}_{j}
        \label{eq:Ham_dis_1bar}
\end{equation}
where $\Delta_j$ is randomly extracted in the interval $[-\epsilon,+\epsilon]$ for each $j$ except for $j_0=\ceil{N_r/4}$ in which $\Delta_j$ belongs to the interval $[\Delta-\epsilon,\Delta+\epsilon]$. Therefore, it corresponds to the case of the single localized barrier in presence of noise. We study the effect of disorder on the leads population for $N_{rea}=100$ disorder realizations in the single barrier case for $N=12$ and $\alpha=3$. We report the behavior of the expectation value
\begin{equation}
\overline{\braket{E_n|\hat{\sigma}_{\mathcal{L}}^z|E_n}}=\dfrac{1}{N_{rea}}\sum_{\beta=1}^{N_{rea}}\braket{E_n^{(\beta)}|\hat{\sigma}_{\mathcal{L}}^z|E_n^{(\beta)}}
\label{eq:Magnetz}
\end{equation}
which is the leads magnetization expectation value in the Hamiltonian spectrum averaged over many disorder realizations.
$\ket{E_n^{(\beta)}}$ is the $n$-th eigenstate of the Hamiltonian for the $\beta$-th disorder realization. Fig.\ref{eq:Ham_dis_1bar} reports the behavior of the leads magnetization for different values of the disorder. We can observe how the presence of disorder can be dangerous for the population; indeed, for strong disorder ($\epsilon=1$) ring and leads states can become $\ket{E_{n_{lr}}}$ states. However, a little amount of disorder ($\epsilon=0.1$) does not modify the structure of the populations. In the intermediate case $\epsilon=0.5$ the population behavior is slightly modified but keeps well the zero disorder structure with a resonance and a well defined separation between leads and ring states far from it.

\section{Phase difference dynamics}
\label{app:phase_diff}
Here we provide more details on the phase effects that the presence of localized detunings generate. In the first part we show how the presence of a single localized detuning in the center of the chain of three atoms has significant effects on the phase associated to each site. For this reason, we suspect that a phase difference between two different paths with detuning located in different positions arises. Thus, in the second part of the appendix and in the main text we report the phase difference between two sites in the two arms in different conditions. We show that its dynamics is related to the population one.

\subsection{Detuning-dependent phase difference in the three sites system}
The presence of a localized detuning introduces a time dependent phase between different sites of the system. To see this, let us consider the simple three sites problem. We suppose to consider only nearest-neighbor hopping, and put a localized detuning in the middle of the chain. As a first stage, we fix the detuning $\Delta=0$ and consider the three site Hamiltonian
\begin{equation}
    \hat{\mathcal{H}}_{\rm hop}=J\left(\hat{\sigma}^+_1\hat{\sigma}^-_2 + \hat{\sigma}^+_2\hat{\sigma}^-_3 + \rm H.c.\right). 
\end{equation}
Supposing to consider the initial state $\ket{\psi(0)}=\ket{\uparrow_1,\downarrow_2,\downarrow_3}$ and evolve it through the Schrödinger equation, the resulting time-evolved state is
\begin{widetext}
\begin{equation}
    \ket{\psi(t)}=\dfrac{1}{2} \Big[ \cos \big( \sqrt{2J}t \big) +1 \Big] \ket{\uparrow_1,\downarrow_2,\downarrow_3} + \dfrac{i}{\sqrt{2}}\sin \big( \sqrt{2J}t \big) \ket{\downarrow_1,\uparrow_2,\downarrow_3} + \dfrac{1}{2} \Big[ \cos \big( \sqrt{2J}t \big) -1 \Big] \ket{\downarrow_1,\downarrow_2,\uparrow_3};
\end{equation}
\end{widetext}
in absence of detuning, there is not a relative phase between the sites $1$ and $3$. In presence of a localized detuning, the Hamiltonian considered is $\mathcal{\hat{H}}=\hat{\mathcal{H}}_{\rm hop}+\Delta \hat{n}_2$, the resulting time-evolved state is
\begin{widetext}
\begin{align}
    \ket{\psi(t)}=&\left(\dfrac{e^{-i(\Delta-f(\Delta,J))t/2}}{\mathcal{N}_1^2} + \dfrac{e^{-i(\Delta+f(\Delta,J))t/2}}{\mathcal{N}_2^2}+\dfrac{1}{2}\right)\ket{\uparrow_1,\downarrow_2,\downarrow_3}+\nonumber\\
    & \left(\dfrac{(\Delta-f(\Delta,J))e^{-i(\Delta-f(\Delta,J))t/2}}{2J\mathcal{N}_1^2} + \dfrac{(\Delta+f(\Delta,J))e^{-i(\Delta+f(\Delta,J))t/2}}{2J\mathcal{N}_2^2}\right)\ket{\downarrow_1,\uparrow_2,\downarrow_3}+ \nonumber\\
    &
    \left(\dfrac{e^{-i(\Delta-f(\Delta,J))t/2}}{\mathcal{N}_1^2} + \dfrac{e^{-i(\Delta+f(\Delta,J))t/2}}{\mathcal{N}_2^2}-\dfrac{1}{2}\right)\ket{\downarrow_1,\downarrow_2,\uparrow_3}
\end{align}
\end{widetext}
where $\mathcal{N}_1^2=2+(\Delta-f(\Delta,J))^2/4J^2$, $\mathcal{N}_2^2=2+(\Delta+f(\Delta,J))^2/4J^2$ and $f(\Delta,J)=\sqrt{\Delta^2+8J^2}$. In this way, the real and imaginary part of the $\ket{\uparrow_1,\downarrow_2,\downarrow_3}$ state are
\begin{subequations}
\begin{align}
    &\Re\left(c_{\uparrow\downarrow\downarrow}(t) \right)=\dfrac{\cos\left((\Delta-f)t/2\right)}{\mathcal{N}_1^2}+\dfrac{\cos\left((\Delta+f)t/2\right)}{\mathcal{N}_2^2}+\dfrac{1}{2},\\&\Im\left(c_{\uparrow\downarrow\downarrow}(t) \right)=-\dfrac{\sin\left((\Delta-f)t/2\right)}{\mathcal{N}_1^2}-\dfrac{\sin\left((\Delta+f)t/2\right)}{\mathcal{N}_2^2}.
\end{align}
\end{subequations}
The phase associated to the state is
\begin{equation}
    \phi_{\uparrow\downarrow\downarrow}(t)=\tan^{-1}\left[\dfrac{\Im\left(c_{\uparrow\downarrow\downarrow}(t) \right)}{\Re\left(c_{\uparrow\downarrow\downarrow}(t) \right)}\right],
\end{equation}
while the phase associated to the state $\ket{\downarrow_1,\downarrow_2,\uparrow_3}$ is 
\begin{equation}
    \phi_{\downarrow\downarrow\uparrow}(t)=\tan^{-1}\left[\dfrac{\Im\left(c_{\uparrow\downarrow\downarrow}(t) \right)}{\Re\left(c_{\uparrow\downarrow\downarrow}(t) \right)-1}\right],
\end{equation}
therefore, the two phases are different. In presence of detuning, a non-zero relative phase between the first and the last site of the chain is present. The difference with the zero detuning case is clear: the states $\ket{\uparrow,\downarrow,\downarrow}$ and $\ket{\downarrow,\downarrow,\uparrow}$ acquire a $\Delta$-dependent phase, also the state $\ket{\downarrow,\uparrow,\downarrow}$ has a $\Delta$-dependent phase differently from the zero detuning case in which it is fixed to $\pm \pi$. 

\subsection{Phase difference for different locations of detunings}
\begin{figure}[!t]
\centering
\includegraphics[width=\columnwidth]{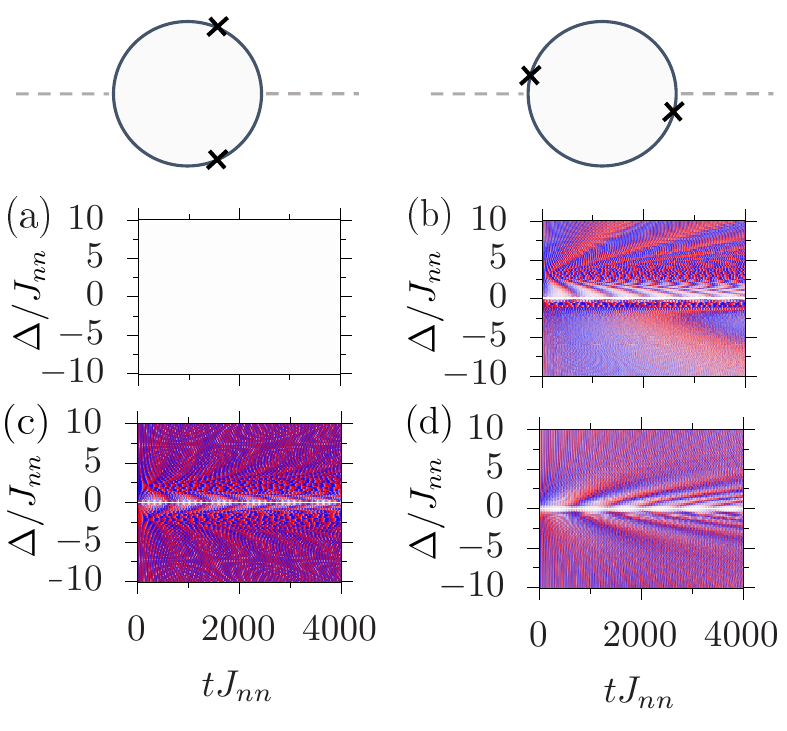}
\caption{Phase difference $\delta\phi$ in function of detuning and time. Panels (a,c): phase difference for two equal (a) and opposite (c) detunings located at $j_A=\ceil{N_r/4}$ and $j_B=\ceil{3N_r/4}-1$ as in the top left sketch. Panels (b,d): phase difference for two equal (b) and opposite (d) detunings located at $j_A=1$ and $j_B=N_r/2+1$ as in the top right sketch. The hopping range is fixed to $\alpha=3$, $K_{nn}/J_{nn}=0.1$}
\label{fig:phase_plot}
\end{figure}
If we consider two arms that differ from the detuning value, we expect to see a non-trivial phase difference between the sites of them. To formally evaluate the phase shift in the set-up considered in the main text, we write a generic time evolved states of the system in the position basis
\begin{widetext}
\begin{equation}
    \ket{\psi(t)}=c_{\mathcal{S}}(t)\ket{\mathcal{S}}+\sum_{j\in \mathcal{R}}c_{j}(t)\ket{\downarrow}_{\mathcal{S}}\otimes\ket{\downarrow,...,\uparrow_{j},...,\downarrow}_{\mathcal{R}}\otimes \ket{\downarrow}_{\mathcal{D}}+c_{\mathcal{D}}(t)\ket{\mathcal{D}},
\end{equation}
\end{widetext}
and we observe that the two body correlator 
\begin{equation}
\braket{\hat{\sigma}_i^{+}\hat{\sigma}_j^-}=c_i^{*}(t)c_j(t)=|c_i(t)||c_j(t)|e^{i(\phi_j(t)-\phi_i(t))}
\end{equation}
is directly related to the phase shift. Thus, the phase difference between the two sites $i$ and $j$ is accessible through $\delta\phi(t)=\phi_j(t)-\phi_i(t)=\arg \big[ \braket{\hat{\sigma}_i^{+}\hat{\sigma}_j^-}\big]$. We can fix $i=N_r/2-1$ and $j=N_r/2+1$ in order to have information on the difference between the phases accumulated by excitation fractions crossing the two arms of the ring. 
Figures~\ref{fig:3dplots_singbarr}, \ref{fig:3dplots_2eqbarr}, and~\ref{fig:3dplots_2opbarr} of the main text report the phase difference dynamics and detuning dependence for the configurations in Fig.~\ref{fig:scheme}, displaying its relation with the population dynamics. Here, we show the phase difference for alternative choices of the detunings location in the Rydberg case $\alpha=3$.

We immediately observe that when the two detunigs have the same sign and their locations are symmetric with respect to the leads [Fig.~\ref{fig:phase_plot}(a)], the phase difference is zero everywhere. Indeed, due to the symmetry of the system, the excitation acquires the same phase in the two arms. Differently, if the detunings are opposite, strong phase difference oscillations appear in such a way that they are never zero except in $\Delta=0$ [Fig.~\ref{fig:phase_plot}(c)]. Figure~\ref{fig:phase_plot}(b,d) show the phase difference in a situation in which the distance between the detunings is the same as the one considered in the main text, meaning $j_B-j_A=N_r/2$, but they are located in different positions. As in the main text case, the equal barrier case presents two resonances for $\Delta_{res}$ analogous to those considered in the main text configuration, they are easily recognizable by fast phase difference oscillations. However, far from resonance, the dynamics is different and strongly depends on the the barrier location. The same applies to the opposite barrier case, there is no resonance but the nature of the oscillation is different.

\section{Strong coupling dynamics}
\label{app:strong_coupling}
In the main text we studied the dynamics in the weak leads-ring coupling regime (in particular, we fixed $K_{nn}/J_{nn}=0.1$). Thanks to this assumption, some important features of the dynamics could be understood through the study of the isolated ring spectrum. If we increase the coupling, the ring and leads modes are no longer separated, the internal dynamics in the ring becomes comparable with those of the leads, and its population becomes significantly non-zero (see \ref{app:strong_spectrum}).

\begin{figure*}[!t]
\centering
\includegraphics[width=0.85\textwidth]{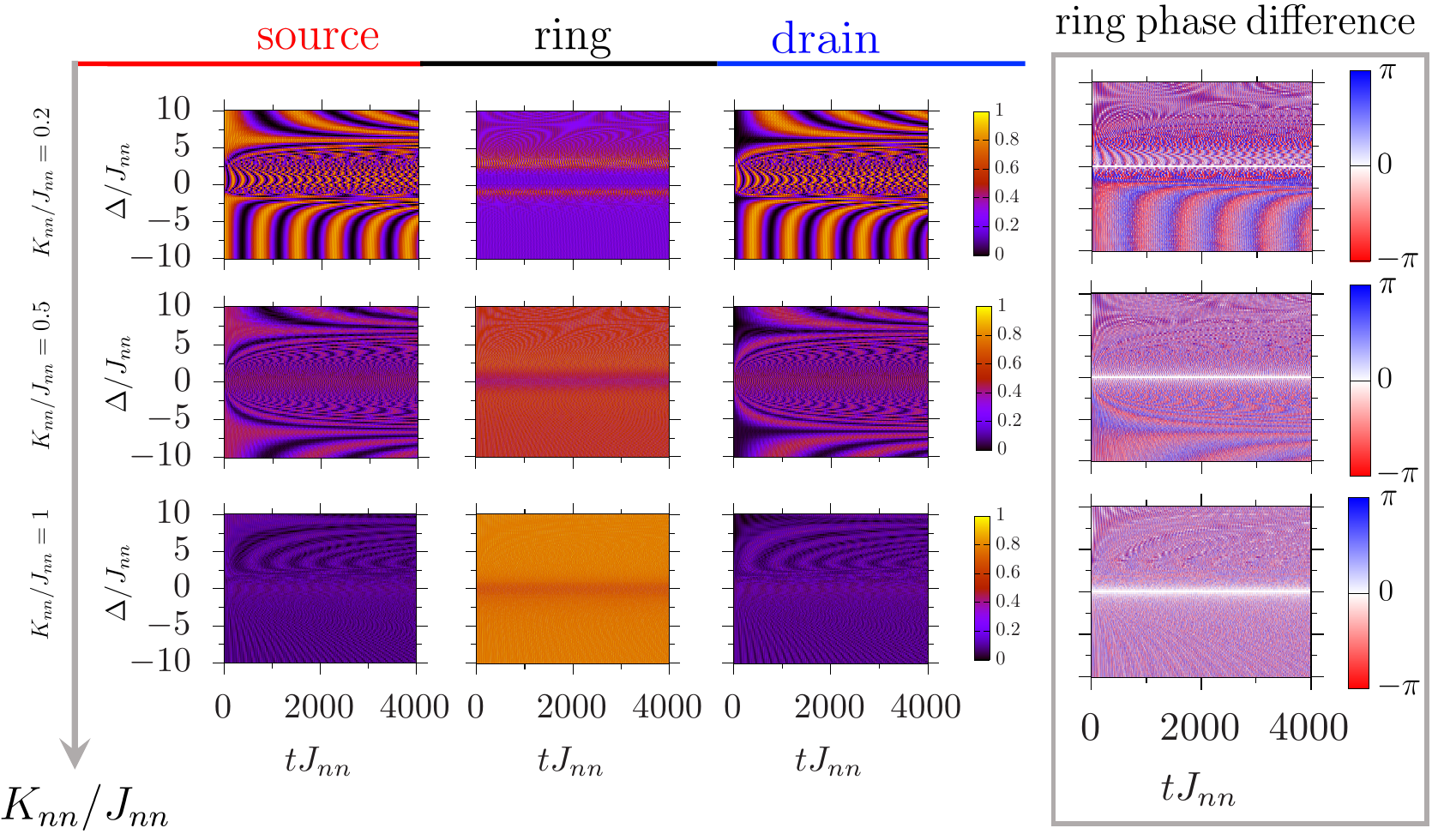}
\caption{Panels on center-left: excitation populations in the source $P_{\mathcal{S}}$, in the ring $P_{\mathcal{R}}$ and in the drain $P_{\mathcal{D}}$ for three different values of the leads-ring coupling $K_{nn}/J_{nn}=0.2,0.5,1$ from top to bottom. Three panels on the right: ring phase difference $\delta\phi$ between two opposite sites located in the two arms of the ring for the three values of $K_{nn}/J_{nn}$ considered. Excitation populations and phase difference are plotted in function of time ($x$-axis) and detuning ($y$-axis). Two equal detunings such that $\Delta_A=\Delta_B=\Delta$ located at $j_A=\ceil{N_r/4}$ and $j_B=\ceil{3N_r/4}$ are considered. The population values are indicated by the color bar that goes from $0$ to $1$, the phase difference by the one that goes from $-\pi$ to $\pi$. The number of sites in the ring is $N_r=14$. The inverse hopping range is $\alpha=3$.}
\label{fig:strong_coupling}
\end{figure*}
Figure~\ref{fig:strong_coupling} reports the population and phase difference behavior for $\alpha=3$, in the presence of equal detunings for different values of the leads-ring coupling. Comparing these results with those in Fig.~\ref{fig:3dplots_2eqbarr} obtained for $K_{nn}/J_{nn}=0.1$, we can recognize clear differences with respect to the weak coupling case reported in the main text. First, the ring presents non negligible population also for values that do not correspond to resonance between ring eigenstates and the source initial state. As regards the source-drain transport, its velocity is increased far from resonance; between the two resonances the oscillations lose regularity due to the presence of ring modes that enter the dynamics. Increasing the coupling, the ring becomes more and more populated and the source-drain oscillations are less evident. For $K_{nn}/J_{nn}=0.5$, independently of the detuning, the spectrum is not anymore separated in leads and ring states, thus the resonance plays a marginal role. The dynamics is characterized by weak source-drain oscillations with a pattern that is almost symmetric for $\Delta \to -\Delta$ and independent on $\Delta_{res}$. Finally, for $K_{nn}/J_{nn}=1$, the ring and leads dynamics are exactly comparable, the ring is almost fully populated and the leads are characterized by weak oscillations.

The rightmost panels of Fig.~\ref{fig:strong_coupling} show the phase difference $\delta\phi$ between the sites located on the two arms of the ring. As in the weak coupling case, $\delta\phi$ follows the population oscillations for different values of $K_{nn}/J_{nn}$. When $K_{nn}/J_{nn}=1$, the ring is strongly populated. Although the population in the leads is small, the phase oscillations seem to be characterized by the frequency of the source-drain population's oscillations. 

Both for the population and the phase difference dynamics, the weak coupling regime turns out to be the most interesting one in terms of controllability of the source-drain excitation transport through the localized detunings placed in the ring.

\section{Linear-chain network}
\label{app:chain}

Here we consider an alternative arrangement of the atoms, with respect to that considered in the main text, in which the channel is made of a linear chain of $N_c=N-2$ atoms. Distances are defined as follows:
\begin{equation}
    \begin{split}
    &d_{i,j}=a|i-j|,
    \quad
d_{\mathcal{S},\mathcal{D}}=d_{\mathcal{S},\mathcal{C}}+(N_c-1)a+d_{\mathcal{D},\mathcal{C}},
\\&    d_{\mathcal{S},j}=d_{\mathcal{S},\mathcal{C}}+a(j-1),
    \quad
    d_{\mathcal{D},j}=d_{\mathcal{D},\mathcal{C}}+a(N_c-j),
    \end{split}
\end{equation}
where $a$ is the lattice spacing in the chain, while $d_{\mathcal{S},\mathcal{C}}$ and $d_{\mathcal{D},\mathcal{C}}$ are the distances between leads and chain. The sites are labeled in such a way that $j=1$ is the site connected to the source and $j=N_c$ is the site connected to the drain. 

\begin{figure*}[!t]
\centering
\includegraphics[width=0.75\textwidth]{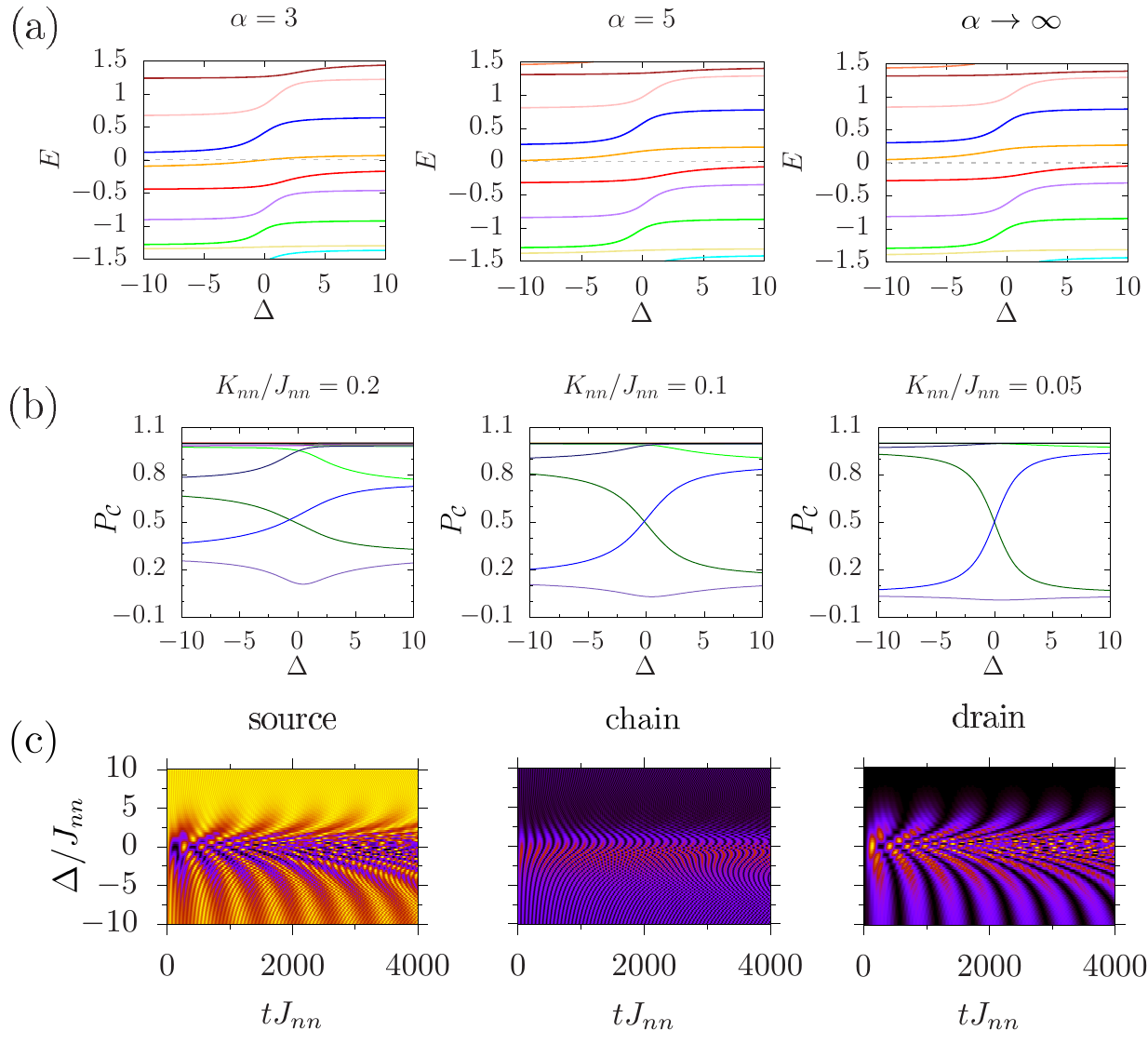}
\caption{(a) Uncoupled spectrum close to zero energy, from left to right: $\alpha=3$, $\alpha=5$, $\alpha\rightarrow\infty$ which is the nearest-neighbor limit; $J_{nn}=1$ and $N_c=14$. (b) Chain populations $P_{\mathcal{C}}$ in the complete spectrum for different values of the ring-leads coupling $K_{nn}/J_{nn}$; $\alpha=3$, $J_{nn}=1$ and $N_c=14$. (c) Dynamics of the source, chain and drain populations (from left to right) for different values of the detuning; the leads-chain coupling is weak $K_{nn}=0.05$, $J_{nn}=1$, the inverse hopping range and the number of atoms are $\alpha=3$ and $N_c=14$. In all the three panels a single detuning located at $j_0=\ceil{N_c/4}$ is considered.}
\label{fig:chain_plot}
\end{figure*}

In the nearest-neighbor coupling limit, the channel Hamiltonian 
can be diagonalized passing first through a Jordan Wigner transformation and then appropriately rotating the creation and destruction operators. The unperturbed Hamiltonian turns out to be~\cite{zawadzki2017symmetries,hegde2015quench}
\begin{equation}
    \hat{\mathcal{H}}_0 =  \sum_{n=1}^{N_c}E_n^{(0)}\hat{d}_n^{\dagger}\hat{d}_n,
    \qquad
    E_n^{(0)} = 2J \cos\left(\dfrac{\pi n}{N_c+1}\right), 
\end{equation}
where the $d^{(\dagger)}_n$ fermions are superpositions of the $c^{(\dagger)}_n$ fermions~\eqref{eq:c-fermions}:
\begin{equation}
    \hat{c}_j=\sqrt{\dfrac{2}{N_c+1}}\sum_{n=1}^{N_c}\sin\left(\dfrac{\pi j n}{N_c+1}\right) \hat{d}_n.
\end{equation}
In this case the spectrum is not degenerate, thus the barrier will simply shift the energy levels without splitting any degeneracy. In the case of non-degenerate spectrum the first order correction in presence of a single detuning located at the position $j_0$ can be straightforwardly calculated
\begin{equation}
   \Delta E_n^{(1)}=\Delta\braket{n|\hat{n}_{j_0}|n}=\dfrac{2\Delta}{N_c+1}\sin^2\left(\dfrac{\pi n j_0}{N_c+1} \right).
\end{equation}
We are interested in the role of the long-range hopping. Fig.~\ref{fig:chain_plot}(a) reports the energy spectrum of the uncoupled chain Hamiltonian near to the zero energy value for different values of $\alpha$. For $\Delta=0$ and in the nearest-neighbor case, the spectrum is symmetric with respect to $E=0$, then a linear shift of the levels is caused by $\Delta$. As in the ring case, with a single detuning the shift is not sufficient to have a resonance at zero energy in the detuning interval considered. Increasing the hopping range and so decreasing $\alpha$, the levels are basically shifted downwards in terms of energy. For this reason, for finite $\alpha$ it is possible that resonances appear. For $\alpha=3$, we observe a resonance with the zero energy state at $\Delta$ around zero. The presence of a resonance will allow us to state that the ring will be populated during the dynamics.

However, as observed previously, the presence of resonances is not sufficient to obtain complete information on the dynamics. It is also important to understand how the complete spectrum of leads and chain behaves in $\Delta$. In Fig.~\ref{fig:chain_plot}(b) we show the chain population for three different values of the leads-chain coupling $K_{nn}/J_{nn}$. We can immediately observe that in this case also for $K_{nn}/J_{nn}=0.1$ chain and leads states are not well separated. In particular, far from resonance, many eigenstates are of the form $\ket{E_{nl}}$ and so the dynamics will not result in coherent oscillations between source and drain. For $K_{nn}/J_{nn}=0.05$ the separation is better defined for big values of $\Delta$, while around the resonance the chain population is $1/2$. Thus, we can conclude that to decouple leads and chain we need a bigger distances between them with respect to ring case.

Finally, Fig.~\ref{fig:chain_plot}(c) reports the dynamics of the source, chain and drain population for $K_{nn}/J_{nn}=0.05$. In proximity of the resonance the chain is heavily populated to the disadvantage of the source. Increasing the value of the detuning and going far from resonance, the transport through the chain is reduced; however, the $\Delta$-window in which the chain is populated and is not simply a narrow region as in the ring case.

%

\end{document}